# Decoupling dislocation multiplication and velocity effects in metals at extreme strain rates

Daniyar Syrlybayev[1], Lavanya Raman[1], Niraj Pramod Atale[2], Bhanugoban Maheswaran[1], Siddhartha Pathak[2], Curt A. Bronkhorst[1], Ramathasan Thevamaran[1,*]

[1]*Department of Mechanical Engineering, University of Wisconsin-Madison, Madison, WI 53706*

[2]*Department of Materials Science and Engineering, Iowa State University, Ames, IA 50011, USA*

* Corresponding author: thevamaran@wisc.edu (RT)

## Abstract

The dynamic behavior of metals is governed by collective dislocation motion and interactions that strongly depend on the applied strain rate. Metals exhibit weak strain rate sensitivity (SRS) below a certain threshold, followed by a distinct SRS upturn at higher loading rates. While this upturn is typically attributed to increased glide resistance at high dislocation velocity due to mechanisms such as phonon drag, the role of strain-rate-dependent dislocation multiplication and microstructural evolution under these extreme conditions remains elusive. Here, we decouple these two strengthening effects and show that, while dislocation velocity primarily governs the SRS upturn, the hardening due to microstructure evolution depends strongly on the initial dislocation density. Our investigation of hardness evolution across ten decades of strain rates in a quenched and tempered martensitic low-carbon steel (LCS) using laser-induced projectile impact tests (LIPIT) and nanoindentation reveals SRS upturn at $\sim 10^7 \ s^{-1}$. By performing in situ re-indentation of the formed craters, we probe the contribution of dislocations generated during initial deformation at different strain rates. We show that while dislocation multiplication plays a negligible role in fine-grained LCS with high dislocation density, a pronounced dislocation multiplication contributes to the hardness increase in pure iron with lower initial dislocation density. Our results show that, depending on the initial microstructure of metals, dislocation multiplication significantly governs high-strain-rate plasticity, in addition to dislocation velocity effects.

## 1. Introduction

Metals used as structural elements and outer shields in automobiles, airplanes, spacecraft, and satellites that bear loads over a wide range of timescales—from static structural loads to micrometeorite impacts as fast as a few kilometers per second—must withstand fatigue and circumvent catastrophic failure [1,2]. Additionally, numerous manufacturing processes, such as forging, cold spray, and shot peening, rely on dynamic deformation of metals to refine microstructures, generate defects, and relieve stress [3–5]. Consequently, a clear understanding of the plastic behavior of metals across strain rates is crucial for advancing these technologies and achieving reliable performance in transportation, aerospace, and military applications [6,7].

Plasticity in metals is controlled by the ubiquitous one-dimensional defects known as dislocations [8–10], which determines the material's hardness evolution with strain, temperature, and strain rate [11,12]. At low to moderate strain rates, dislocation movement is thermally activated, resulting in a relatively weak logarithmic strain-rate sensitivity [12,13]. The origins of the material's strain-rate-sensitive plastic response under these conditions are typically attributed to strain-rate-dependent structural evolution as the result of dislocation multiplication and rate-dependent resistance against the glide as the average dislocation velocity increases [12]. It is generally accepted that the strain-rate sensitivity of face-centered cubic (FCC) metals is governed by rate-dependent microstructural evolution due to their low lattice friction, whereas in body-centered cubic (BCC) metals, it is controlled by dislocation glide velocity because screw dislocations encounter a much higher Peierls barrier [12,14].

On the other hand, recent studies suggest a distinct strain-rate sensitivity (SRS) upturn above a strain-rate threshold of around $10^5 - 10^8 \ s^{-1}$, which has been observed in a variety of materials and across different deformation modes, including BCC [15,16] and FCC [17–19] metals subjected to microscale ballistic impact [19–22] and macroscale dynamic compression experiments [15,16,23]. It is believed that the upturn marks a transition from thermally-activated glide, where the resolved shear stress (RSS) remains below the Peierls barrier [24–26], to a phonon-drag-dominated regime, in which the RSS exceeds this barrier. At this regime, the rapid dislocation glide becomes increasingly damped by lattice vibrations (phonons), setting a much stronger linear dependence of the flow stress and hardness on the strain-rate [10,13,27–29]. Observations of thermal hardening at higher strain rates due to increased phonon density support this interpretation [24,30,31]. Another mechanism, rough slip, has also been observed in molecular dynamics (MD) simulations of BCC iron [32]. It is a high-strain-rate deformation mechanism driven by the simultaneous formation of kink pairs on multiple slip planes along the single screw dislocation line. These kink pairs propagate along the dislocation line and form cross-kinks spanning multiple slip planes with sessile segments, causing screw dislocations' self-pinning.

While phonon drag and rough slip occur as the average velocity increases, the influence of rate-dependent structural evolution and potential inertial effects at high-strain-rate loading remains unclear. Post-mortem analysis has shown that a rise in strain rate increases the defect density in Al [14,21,33,34], Ag [35–37], and Cu [38]. The structural evolution in Ag led to the formation of nanograins and substantially improved mechanical properties after deformation [36,37,39]. Similarly, reloading of Cu after high-strain-rate deformation showed an increase in flow stress that is not attainable by quasi-static deformation [40,41].

Both velocity-induced resistance to dislocation glide and defect generation likely contribute to the hardening observed beyond the SRS upturn; yet their respective roles have been examined only scarcely in prior work [14,40,42]. This lack of clarity is particularly pronounced for BCC metals, where reports of an SRS upturn are far less common than in FCC systems. Yet BCC metals—especially iron-based alloys—remain the most widely used structural materials [43].

Therefore, in this study, we examine the origins of the SRS upturn in two related BCC material systems with distinct initial microstructures: quenched-and-tempered martensitic low-carbon steel (LCS) and pure iron. Using LIPIT and nanoindentation (NI), we investigate the evolution of hardness over ten orders of magnitude in strain rate. A newly developed crater re-indentation experiment—measuring quasi-static hardness within the craters formed during initial indentation—combined with detailed microstructural characterizations, reveals that the SRS upturn in hardness is primarily governed by increased resistance to dislocation motion associated with elevated dislocation velocities. Besides, dislocation multiplication contributes to additional hardening that weakly depends on the applied strain-rate and is strongly influenced by the initial microstructure, thereby affecting the material's ability to change properties after high-speed deformation.

## 2. Materials and Methods
### 2.1. Materials

We sourced quenched-and-tempered martensitic low-carbon steel (LCS) from a 0.5-inch-thick plate (Chapel Steel, USA) and compared its performance with that of 99.95% pure iron (Fischer Scientific, USA). Both samples were mechanically ground using 240, 600, 800, and 1200 grit SiC papers, polished with 1 and 0.25 $\mu m$-diamond suspension, and 0.02 $\mu m$-colloidal silica (Allied High Tech, USA) to achieve a mirror finish with low residual surface stress for hardness and microstructural characterization.

Fig. 1a shows an inverse pole figure map of LCS steel, indicating its microstructure consists of larger-scale grains (blocks) with high intragranular misorientations, visible as color contrast variation, typical of a lath martensite structure. The microstructure does not display significant texture, as shown in the pole figures in Fig. 1c. The area-weighted mean grain diameter of ~4.64 $\pm$ 2.83 $\mu m$. Fine grains and large intragranular misorientations result in a high flow stress of ~1800 MPa and low strain hardening under compression in a custom-built Kolsky bar at a strain rate of 2500 s$^{-1}$ (details of a setup in SI Section 1), as shown in Fig. 1e. The microstructure differs from that of pure iron, illustrated in Fig. 1b, which consists of large grains (> 50 $\mu m$) and exhibits a stronger texture, as shown in the pole figures (Fig. 1d).

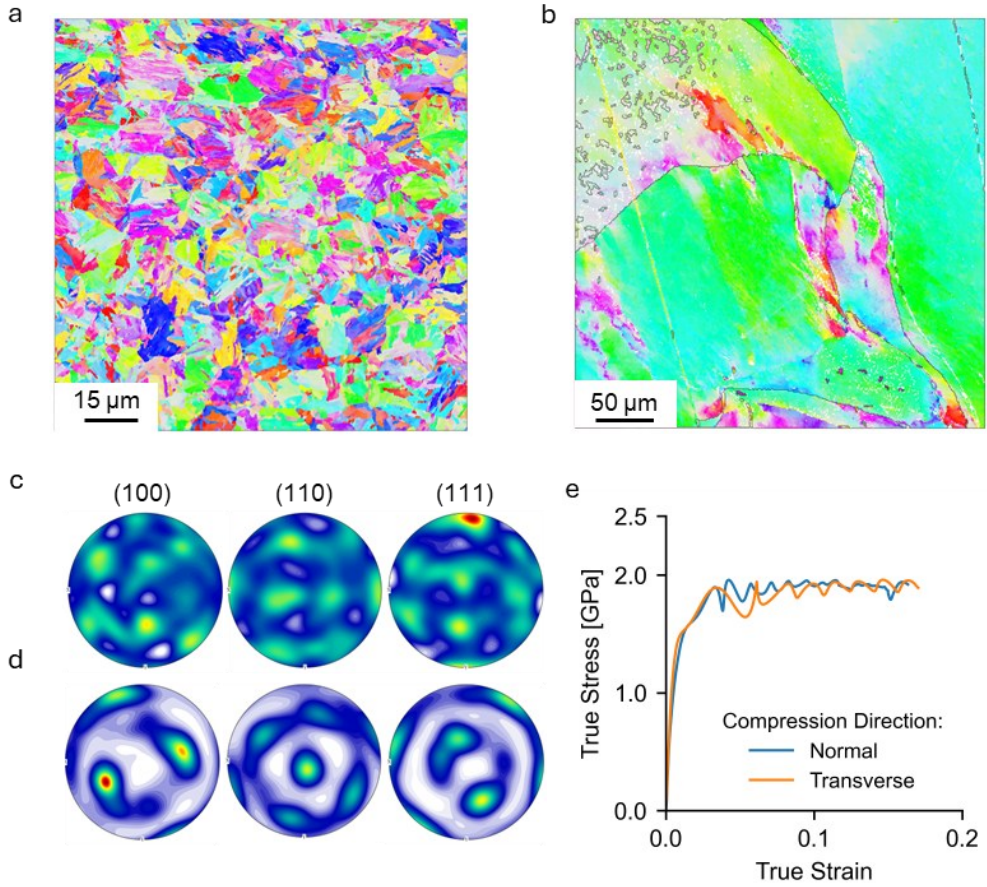

**Fig. 1. Initial material characterization.** (a) IPF map of the LCS material and (b) pure iron; (c) Pole figures of the LCS material and (d) pure iron; (e) True stress–strain response of the LCS material under Kolsky bar compression at 2500 $s^{-1}$.

### 2.2. Pristine surface hardness measurements across strain rates

To study the hardness evolution of the LCS material and pure iron across strain rates, we used nanoindentation (NI) and laser-induced projectile impact testing (LIPIT). Both methods involve driving a spherical indenter into the material at different speeds defining strain-rate. We performed low and medium-strain-rate spherical nanoindentation tests on the LCS sample using a 20 $\mu m$ radius spherical indenter (manufactured by Synton-MDP, Switzerland) to a maximum depth of ~1800 nm with an *in-situ* SEM nanoindenter (Alemnis AG, Switzerland) in the Ultra-High-Strain-rate (UHS) configuration. During these tests, a piezoelectric actuator applies a loading function, given by $\dot{P}/P = const$, to the indenter tip as shown in Fig. 2a. All experiments were carried out in an FEI Inspect 50 SEM (FEI, USA) to ensure a precise positioning of the indenter tip during indentation.

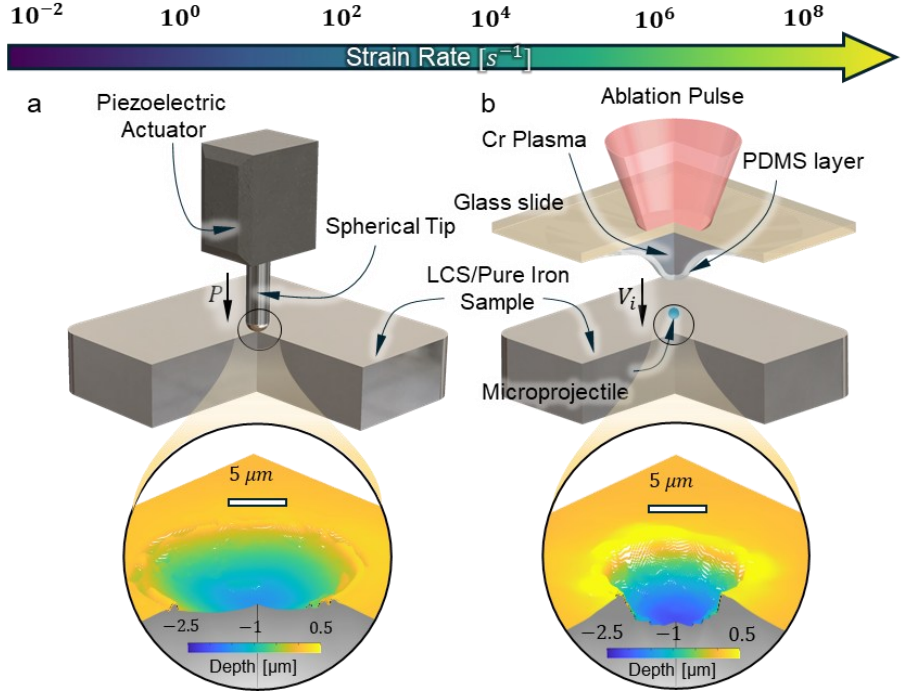

**Fig 2. Hardness across strain rates.** An experimental framework to investigate hardness across ten orders of magnitude in strain rate using (A) quasi-static and dynamic nanoindentation and (B) laser-induced projectile impact testing (LIPIT).

The high-strain-rate response ($\sim 10^6 - 10^7 \, s^{-1}$) was investigated with the custom-built LIPIT setup, as shown in Fig. 2b, with detailed information of the setup available in our previous work [21]. We launched $30 \pm 2.1 \, \mu m$ and $12 \pm 2.3 \, \mu m$ diameter nearly-spherical alumina projectiles—hereafter referred to as the 30 $\mu m$ and 12 $\mu m$ projectiles, respectively—at velocities from 200 to 700 $m \, s^{-1}$ onto the target material. Impact and rebound velocities were recorded using a high-speed camera system, and position and velocity histories of each particle during impact were tracked as described in SI Section 2. After deformation, we measured the topography and volume of each crater and its surroundings using a white-light interferometer (Zygo NewView 9000, USA) to evaluate its geometrical characteristics for hardness calculations. Representative examples of the measured topography are illustrated in the insets of Fig. 2.

*2.3. Finite Element Modelling*

To compare the material and mechanical response under LIPIT and NI, we modeled the deformation using finite-element simulations. For this purpose, we calibrated the Johnson–Cook plasticity model [44] for the LCS material, which is given by

$$\sigma = (\sigma_y + B\varepsilon_p^n)(1 + C\ln\frac{\dot{\varepsilon}_p}{\dot{\varepsilon}_r})(1 - \hat{\theta}^m) \qquad (1)$$

$$\hat{\theta} = \begin{cases} 0 \text{ if } T < T_r \\ \dfrac{T - T_r}{T_m - T_r} \text{ if } T_r < T < T_{melt} \\ 1 \text{ if } T \geq T_{melt} \end{cases}$$

with $\sigma_y, B, n, C, \dot{\varepsilon}_r, m$ being material constants, $\sigma, \varepsilon_p$ and $\dot{\varepsilon}_p$ are flow stress, plastic strain, and plastic strain-rate, respectively, and $\theta$ is a function of temperature $T$, $T_r$ is reference temperature (298 K), and $T_{melt}$ melting temperature of 1793 K for steel. Furthermore, we assumed an isotropic, linear, and temperature-independent material elastic response with a Young's modulus of 210 GPa and a Poisson's ratio of 0.3.

The simulation considered deformation under NI at strain rates of ~0.05, ~5, and ~270 s$^{-1}$, along with 30 $\mu m$ and 12 $\mu m$ projectile LIPIT impacts. The deformation under NI and 30 $\mu m$ impact cases was used for model calibration, whereas the 12 $\mu m$ impact simulations were employed for verification. Axisymmetric deformation was assumed, and the simulation domain was chosen to be at least ten times the indenter tip radius.

We calibrated the model using nanoindentation load-displacement data with the FEM model setup shown in Fig. S6a. A displacement history $h(t)$ obtained directly from the nanoindentation experiments as shown in Fig. 3a was applied to the rigid indenter with the 20 $\mu m$ radius and 45° tip half-angle. To ensure numerical stability under large deformations and to reduce mesh distortion, the near-field region was discretized with CAX3 elements, while the far-field region used CAX4R elements (Fig. S6a). Mesh convergence was tested by gradually decreasing the near-field element size from 1 $\mu m$ to 0.25 $\mu m$ (Fig. S6c). Additional simulations using CAX4R elements in the near-field region showed no significant deviation from the triangular-mesh results, as shown in Fig. S6c. Based on these analyses, we selected the 0.5 $\mu m$ CAX3 element size for nanoindentation (1/40 of the indenter tip radius). This mesh-to-tip-radius ratio was kept at least at this level in the subsequent LIPIT simulations, using 0.5 $\mu m$ elements for the 30 $\mu m$ projectile and 0.1 $\mu m$ elements for the 12 $\mu m$ projectile.

For calibration, we first considered two nanoindentation cases with nominal strain rates of 0.05 and 270 $s^{-1}$. Temperature effects are negligible in the low–strain-rate case (0.05 $s^{-1}$), because the process is slow and isothermal. For the higher strain rate (~270 $s^{-1}$), the temperature rise was calculated from the relation

$$\rho C_v \dot{T} = \beta\, \boldsymbol{\sigma}:\dot{\boldsymbol{\varepsilon}} \qquad (2)$$

[44] where $\rho = 7870\ kg\ m^{-3}$ is the density of steel, $C_v = 466\ J\ kg^{-1} K^{-1}$ is the heat capacity at constant volume, $\beta = 0.9$ is the Taylor-Quinney factor, $\boldsymbol{\sigma}$ and $\dot{\boldsymbol{\varepsilon}}$ are the stress and strain rate tensors. Results showed that the expected temperature increase would be approximately 50 K at a plastic strain of ~0.22. Due to a relatively small increment, we ignored thermal effects in NI simulations and calibrated material parameters $\sigma_y, B, n, C, \dot{\varepsilon}_r$.

We used a custom Python optimization algorithm to fit the model. The algorithm accepts the Johnson–Cook parameters $\sigma_y, B, n, C, \dot{\varepsilon}_r$ as input, feeds them into a model with the specified boundary condition and deformation history $h(t)$, and then retrieves the resulting simulated load–displacement response $P_{sim}$, which is compared to the experimental result $P_{exp}$. The objective function used in the optimization is defined by

$$L = \frac{1}{n}\sum \frac{(P_{sim} - P_{exp})^2}{P_{exp}} \tag{3}$$

This loss function prioritizes larger loads and their associated indentation depths because local crystallographic orientation effects may affect the response at shallow depths, given the small sampled volume. As indentation depth increases, the response approaches bulk behavior and is therefore more suitable for calibration. The Nelder–Mead algorithm was selected for minimizing the objective function.

Optimization yielded the model parameters $\sigma_y, B, n, C, \dot{\varepsilon}_r$ that minimize the objective function $L$. To determine the thermal softening parameter $m$ we used the LIPIT results obtained with a 30 $\mu m$ diameter projectile. Due to the short impact duration, all LIPIT simulations were performed using an explicit solver, while the overall model setup remained similar to that of NI, as shown in Fig. S6b. The parameter $m$ was optimized such that the simulated residual crater depth matched the experimentally measured depth at the corresponding impact velocity. Table 1 summarizes the material constants obtained. The resulting model was subsequently verified using the 12 $\mu m$ projectile impact.

**Table S1.** Summary of material parameters used in the model

| Constant | Values | Constant | Values |
|---|---|---|---|
| $E\ (GPa)$ | 210 | $C$ | 0.1413 |
| $\nu$ | 0.3 | $\dot{\varepsilon}_r\ (s^{-1})$ | 0.2 |
| $\sigma_y\ (MPa)$ | 1122.92 | $C_v\ (J/kg \times K)$ | 466 (Ref. [45]) |
| $B\ (MPa)$ | 57.7632 | $\beta$ | 0.9 (Ref. [45]) |
| $n$ | 0.00862 | $\rho\ (kg/m^3)$ | 7870 |

*2.4. Self-consistent definitions for hardness and nominal strain rate*

Because of inherent differences in testing methods and diagnostics between NI and LIPIT, the two techniques use different measures of hardness and nominal strain rate. To properly interpret the evolution of hardness across the broad strain-rate range covered by both methods, we established self-consistent definitions of hardness and nominal strain rate for NI and LIPIT. For this, aligned with dynamic experiments, we defined hardness for both NI and LIPIT as a ratio of the total work ($W_p$) by the volume of the crater ($V_c$) formed. In the case of nanoindentation this can be written as

$$H = \frac{W_p}{V_c} = \frac{\oint P \, dh}{V_c} \qquad (4)$$

Where $P$ is the applied load and $h$ is the indentation depth. Note that Meyer's definition of hardness, commonly used in NI, is expressed as a ratio of the peak load to the projected area $A$ ($\bar{H} = \frac{P}{A}$). Therefore, Eq. 4 can be written as

$$H = \frac{W_p}{V_c} = \frac{\oint \bar{H} A(h) \, dh}{\int A(h) \, dh} \qquad (5)$$

The equation shows that as long as hardness $\bar{H}$ remains independent of the indentation depth $h$ and elastic recovery is small, both Meyer's and energy-based definitions of hardness are equivalent. For the LIPIT impact of a nearly rigid projectile on a deformable substrate, the total work is calculated from energy conservation as the change in the projectile's kinetic energy. For the spherical projectile with mass $m$ with the impact velocity $v_i$ and rebound velocity $v_b$ the hardness is given by

$$H = \frac{m(v_i^2 - v_r^2)}{2V_c} = \frac{4}{3}\pi(R)^3 \rho_p \left(\frac{v_i^2 - v_r^2}{2V_c}\right) \qquad (6)$$

where $\rho_p = 3720 \, kg \, m^{-3}$ is the nominal density of alumina projectiles used in this work and $R$ measured impactor radius.

Similar to previous LIPIT works [24,46] we define the nominal strain rate for both NI and LIPIT as

$$\dot{\varepsilon} = \frac{v_{ave}}{R} \qquad (7)$$

where $v_{ave}$ is the work-averaged velocity. In our NI experiments, the work-averaged indenter velocity was estimated directly from the displacement–time history shown in Fig. 3a. Because displacement does not vary linearly with time, different portions of the loading curve correspond to different instantaneous indenter velocities—and consequently to different incremental work inputs. This is shown in Fig. 3b, where we plot the indenter velocity $\dot{h}(h(t))$ as the function of the cumulative normalized work $\frac{1}{W_p}\int_0^{h(t)} P \, dh$. Under these conditions, simple time-averaging underestimates the nominal strain rate. Therefore, to obtain a representative average velocity, we weighed the instantaneous velocity by the corresponding work input. The resulting work-averaged indenter velocity is estimated by

$$v_{ave} = \frac{1}{W_p}\int_0^{h_{max}} \dot{h} \, dW = \frac{1}{W_p}\int_0^{h_{max}} \dot{h} P \, dh \qquad (8)$$

Because of limited diagnostics in LIPIT and the inability to measure the projectile velocity history during impact, we relied on our FEM results to analyze how the projectile velocity evolves during impact on LCS. The indentation depth history is shown in Fig. 3c for several representative impact scenarios. The impact of the 30 $\mu m$ diameter projectile is longer in duration, rebounding ~15 ns after the first contact, while the 12 $\mu m$ diameter projectile impact duration is ~6 ns. The corresponding indentation velocity as a function of cumulative work is shown in Fig. 3d. The work-averaged velocity, defined according to Eq. 8 as a function of impact velocity, is illustrated in Fig. 3e. As can be seen, the FEM-derived work-averaged velocity scales approximately as $\sim(2/3)\,v_{\text{im}}$ within a given range of impact velocities. As a result, consistent with NI experiments, the nominal strain rate under LIPIT is given by

$$\dot{\varepsilon} = \frac{v_{ave}}{R} = \frac{2v_{im}}{3R} \tag{9}$$

Note that this definition results in a modest difference compared to the convention used in LIPIT, whereby the impact velocity is normalized by diameter ($\frac{v_i}{D} = \frac{v_i}{2R}$) [24,46].

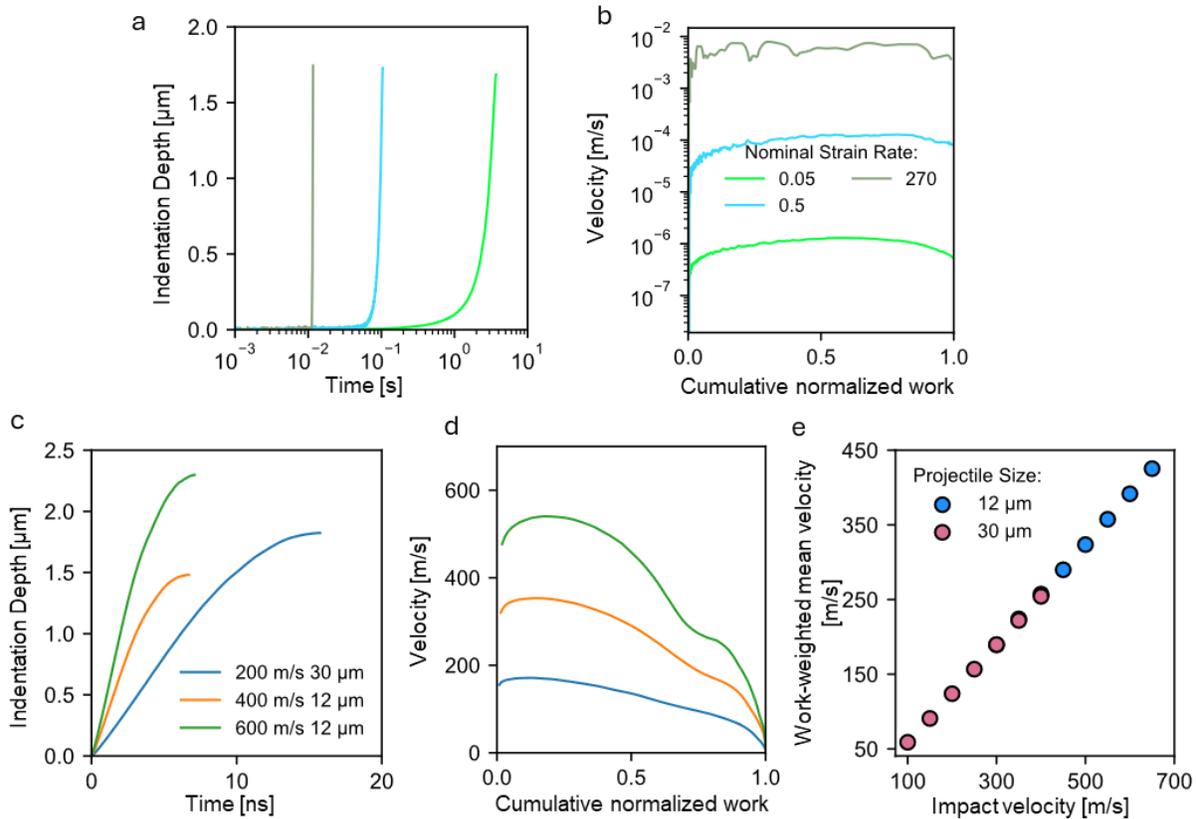

**Fig. 3. Self-consistent hardness and strain rate definition**. (a) Indentation depth versus time during nanoindentation (NI) for several representative test cases. (b) The corresponding indentation velocity as a function of cumulative normalized work. (c) Indentation depth evolution during LIPIT obtained from finite-element simulations for several impact conditions. (d) The corresponding velocity histories as a function of cumulative normalized work. (e) Work-weighted mean indentation velocity plotted as a function of impact velocity.

## 2.5. Crater re-indentation analysis

The initial deformation of the substrate causes microstructural changes—dislocation generation and rearrangement—that increase the material's stored internal energy. To quantify their contribution, we re-indented the craters and measured the quasi-static hardness of the pre-deformed material as shown in Fig. 4a. The conventional Oliver and Pharr analysis for nanoindentation [47] assumes a flat surface, which is not the case with curved craters. Therefore, we modified the conventional protocol to account for changes in contact stiffness and projected area resulting from crater curvature.

In nanoindentation of a flat surface, the reduced modulus $E_r$ can be calculated from the measured unloading compliance $C_e^f$ and the projected area $A$ estimated from the contact depth. The relation is given by:

$$C_e^f = C_{frame} + \frac{\sqrt{\pi}}{2 * E_r \sqrt{A}} \quad \quad 10)$$

where $C_{frame}$ is the frame compliance and superscript $f$ indicates a flat surface measurement. Here, we assume that the crater geometry introduces another uncorrelated linearly elastic compliance source $C_g$ affecting the measurement. Therefore, the measured compliance is a sum of the geometric ($C_g$), frame ($C_{frame}$), and intrinsic material compliances:

$$C_e^c = C_g + C_{frame} + \frac{\sqrt{\pi}}{2 * E_r \sqrt{A}} = C_m + \frac{\sqrt{\pi}}{2 * E_r \sqrt{A}} \quad \quad 11)$$

where superscript $c$ indicates a crater. The two non-material-related compliance sources are combined into the miscellaneous compliance $C_m$. To separate non-material compliances from the total response inside each crater, we impose an oscillating loading function shown in Fig. 4b. The corresponding unloading compliance at each load level ($P$) is calculated from the load-displacement curve (Fig. 4c). Following the method given in Refs. [48,49] we define the quantity $K = C_e^c \sqrt{P}$ which is expressed by:

$$K = C_e^c \sqrt{P} = (C_m)\sqrt{P} + \frac{\sqrt{\pi}\sqrt{P}}{2 * E_r \sqrt{A}} =$$
$$= (C_m)\sqrt{P} + \frac{\sqrt{\pi}\sqrt{H}}{2 * E_r} = (C_m)\sqrt{P} + \frac{\sqrt{J\pi}}{2} \quad \quad 12)$$

In this equation $J = \frac{\sqrt{H}}{E_r}$ is a material-dependent Joslin-Oliver parameter [48,50]. As the equation shows, $K$ is a linear function of $\sqrt{P}$ with the intercept $\frac{\sqrt{J\pi}}{2}$ and the slope $C_m$, which corresponds to the miscellaneous compliance. The dependence of $K$ on $\sqrt{P}$ is illustrated on the SYS plot [48,49] in Fig. 4f. That compliance, which is unrelated to the material characteristics, is subtracted from the measured compliance $C_e^c$ to extract reduced modulus $E_r$ using Eq. 11.

Hardness in re-indentation experiments was calculated as the ratio of the applied force to the projected area of the indenter, which was found from the cube-corner tip area function. To do this, we follow the Oliver and Pharr method [47] and first find an uncorrected contact depth given by:

$$h_c = h_{max} - \epsilon P C_e^c \qquad 13)$$

where $C_e^c$ is the measured compliance of the material, $h_{max}$ – maximum depth, and $\epsilon \approx 0.72$ being the geometric correction factor [47]. Here, we used the measured compliance, assuming that only the indent undergoes permanent deformation, while the deformations of the surrounding material and the frame are elastic, fully recovering during unloading, leaving only the indent-related permanent deformation depth $h_c$. The curvature increases the effective contact depth by an amount $h_c'$, as shown in Fig. 4d, and a refined contact depth estimate $h_c + h_c'$ is obtained from the cube corner geometry using the relation:

$$\frac{7}{4}h_c'^2 + h_c'\left(\frac{3}{2}h_c - 2R_c\right) + \frac{3}{4}h_c^2 = 0 \qquad 14)$$

where $R_c$ is the radius of the crater's curvature (Fig. 4d), determined using a white light interferometer by fitting a spherical profile to each crater cap. Note that at smaller contact depths, $h_c$ and larger curvature radii $R_c$ corrected and uncorrected projected areas converge as shown in Fig. 4e.

This approximation assumes that the indenter tip is slightly offset from the crater center, as shown in Fig. 4d, so that the projected contact area plane remains parallel to the sample surface. As shown in Fig. 4g and summarized in Table S1, the Young's modulus values obtained from these experiments are consistent with the nominal bulk modulus of LCS and pure iron (~210 GPa), thus validating the assumptions used in the analysis. This value can vary over a broader range (~148–245 GPa) depending on the crystallographic orientation within the small interaction volume sampled by a cube-corner tip [51].

The evolution of hardness with indentation depth is shown in Fig. 4h. At shallow indentation depths, the measured hardness is likely affected by the crater's roughness, which introduces frictional forces on the tip without creating deep indents, thereby artificially raising the apparent hardness. With increasing contact depth and applied force, this effect diminishes, and the hardness stabilizes at ~600 nm. At these depths, the indenter remains within the highly deformed region inferred from microanalysis, described further.

The re-indentation experiments were conducted with a Hysitron PI-85R (USA) *in situ* SEM pico-indenter equipped with a cube-corner tip and operated inside a Zeiss 1550 VP Scanning Electron Microscope (SEM, Germany). For LCS, we performed crater re-indentation experiments at each prior strain-rate. This included two to three tests, performed on craters achieved under spherical

NI, as well as five and seven re-indentations of craters created by 30- and 12-$\mu m$-diameter projectile impacts under LIPIT. The 12 $\mu m$ projectile reindentation results for LCS material were compared with those obtained for pure iron under similar prior deformation conditions.

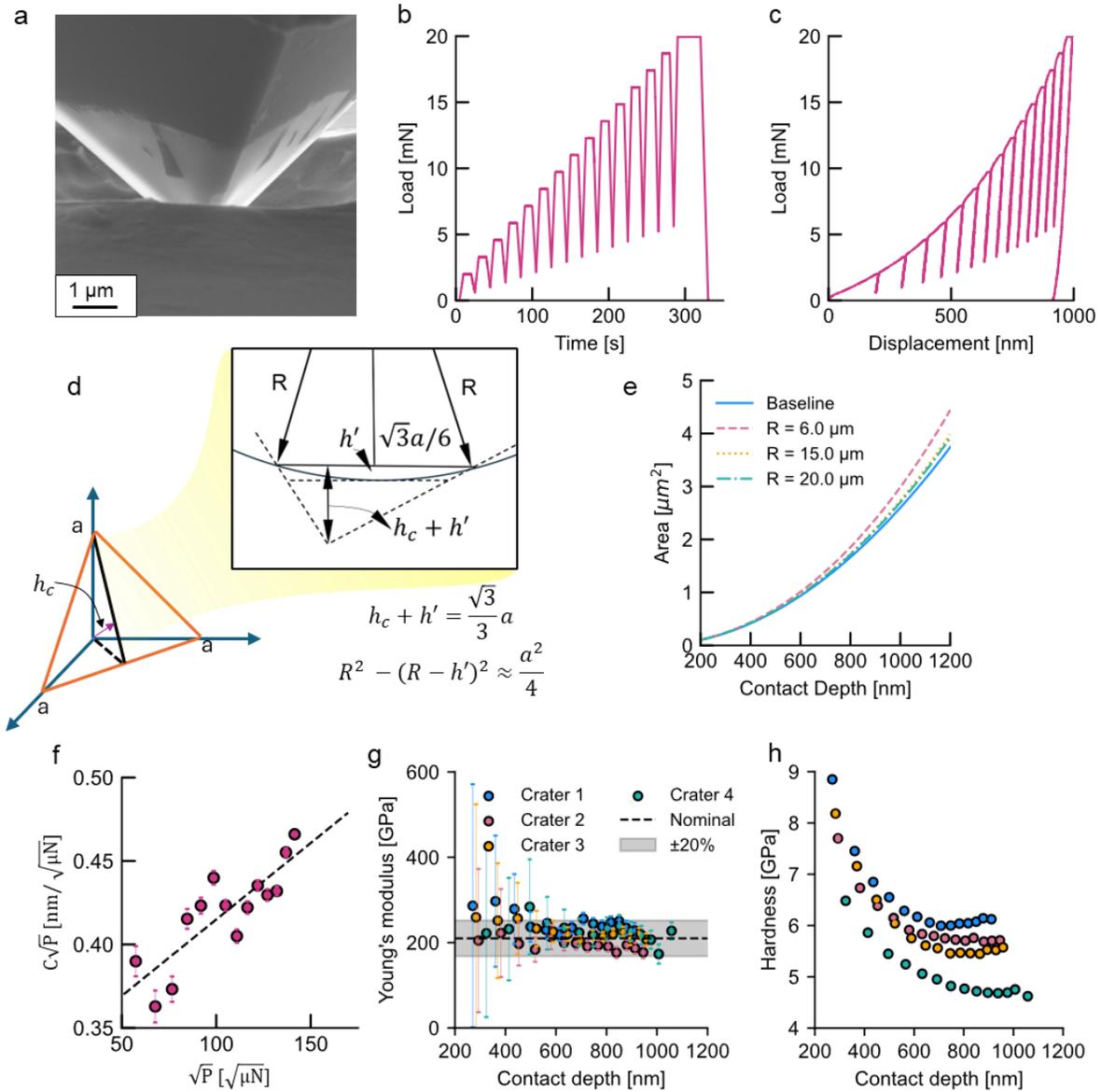

**Fig. 4. Crater Re-indentation Experiments.** (a) SEM image showing the cube-corner indenter positioned inside the impact crater. (b) Applied loading function and (c) the corresponding load-displacement response. (d) Geometric schematic of the cube-corner indenter engaging the curved crater surface. (e) Comparison of corrected contact areas for different crater radii relative to uncorrected values. (f) SYS plot with fitted slope representing compliance arising from the crater geometry and frame. (g) Young's modulus extracted from re-indentation across several representative $12 - \mu m$ impact craters, showing good agreement with the baseline modulus despite irregular surface curvature. (h) Corresponding hardness evolution with indentation depth.

*2.6. Post-mortem microanalysis*

We performed a microanalysis to examine the microstructural changes induced by deformation as a function of strain rate. This involved exfoliating the material beneath the deformed crater using an FEI Helios G4 Xe plasma-focused ion beam (pFIB, USA), followed by EBSD analysis on a Zeiss Gemini 300 SEM (Germany) equipped with an Oxford Symmetry 3 camera (UK). Coarse FIB milling was performed at 30 kV with gradually decreasing currents of 60, 15, and 4 nA. The material was then lifted out and attached to a FIB grid, followed by fine polishing with currents of 1, 0.3, and 0.1 nA. EBSD was conducted at a standard 70° tilt and 20 kV beam with a smaller current, achieved with a 60-$\mu m$ lens aperture and a sampling step size of 25 nm. These settings enabled us to maintain a lower interaction volume, thereby allowing higher spatial resolution. The severe plastic deformation after the impact caused the conventional Hough Indexing (HI) of EBSD patterns in the native Oxford Aztec acquisition software to perform poorly. To address this, we implemented dictionary indexing (DI) as described in Section 3 of the SI. DI improved the accuracy of local orientation measurements in highly deformed regions, achieving better angular accuracy compared to HI. Reindexed local orientations were then processed using custom MATLAB code and the MTEX library to reveal microstructural evolution.

## 3. Results and Discussions

*3.1. Hardness across strain rates*

We first examined the evolution of hardness in LCS across ten orders of magnitude of nominal strain rate, using self-consistent definitions between NI and LIPIT as shown in Fig. 5a. At low to moderate strain rates, deformation is controlled by thermally-activated slip. In this regime, hardness increases logarithmically with strain rate, as shown by nanoindentation results spanning a strain-rate range of ~0.05 to ~270 $s^{-1}$. Extrapolating this trend aligns well with the hardness results calculated from the LIPIT impact of 30 $\mu m$ projectiles at velocities of 200–350 $m\ s^{-1}$, suggesting that thermally-activated slip prevails up to ~$10^7$ $s^{-1}$. At higher strain-rates, achieved by reducing projectile size, we observe a deviation from the initial trend—a clear upturn in SRS (Fig. 5(a, b)).

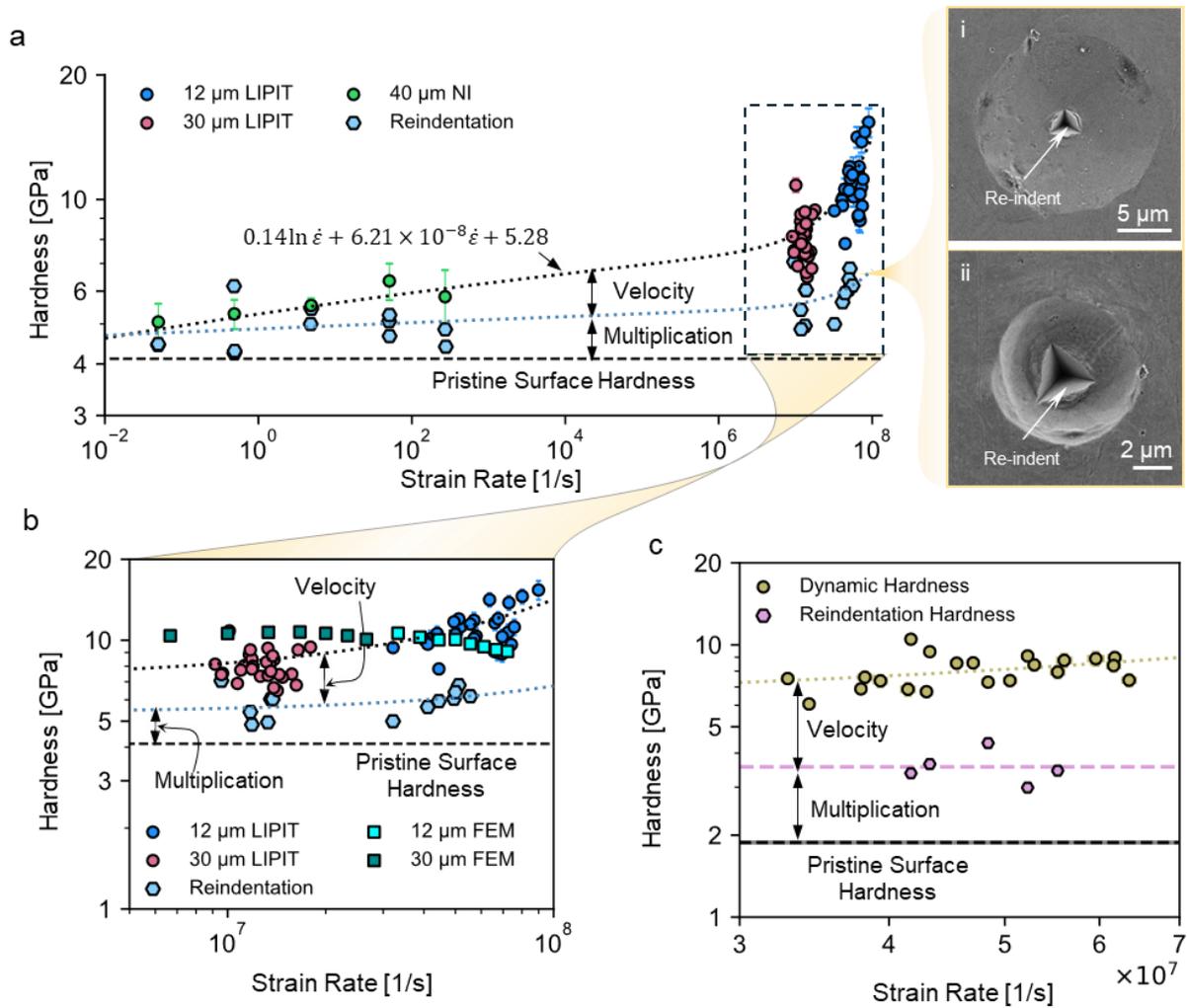

**Fig. 5. Hardness versus strain rate**. (a) The evolution of in-situ and re-indentation hardness of quenched and tempered martensitic Low Carbon Steel (LCS) across the strain-rate spectrum reveals an upturn at $\sim 10^6$–$10^7$ s$^{-1}$. (b) A magnified view of the LCS response under LIPIT, compared with (c) the corresponding response of pure iron.

This SRS upturn has been seen across different loading configurations [22–25] and is typically attributed to a transition from thermally-activated to phonon-drag-controlled dislocation motion. In this regime, lattice vibration (phonon) damps the dislocation mobility [52], causing the strain-rate sensitivity to transition from a logarithmic to linear dependence [42]. This upturn was also previously linked to the appearance of the inertial effects at high rates [53–55] suggesting that, rather than being an inherent material property, it may occur due to mechanical differences between testing methods.

*3.2. Delineating mechanistic differences between LIPIT and NI using FEM*

To analyze the mechanical differences between LIPIT and NI, we used finite-element results. Fig. 6a compares the FEM-predicted load–displacement curves with the nanoindentation measurements, using the calibrated material model, and shows good agreement. Similarly, Fig.

6(b, c) compare the FEM-estimated crater depth as a function of impact velocity with the corresponding LIPIT measurements for 30 $\mu m$ and 12 $\mu m$ projectile, respectively, further validating the material model.

We then evaluate the dynamic hardness from the FEM using Eq. 6, as shown in Fig. 5b. Unlike the experimental results, the FEM-predicted dynamic hardness exhibits slight softening. This softening arises because of the reduction in the projectile size from 30 $\mu m$ impact to 12 $\mu m$ elevates not only the nominal strain-rate, from ~$1.33 \times 10^7$ $s^{-1}$ to ~$7.2 \times 10^7$ $s^{-1}$ at a given impact velocity—but also the temperature, which rises from up to ~600 K at 30 $\mu m$ projectile impacts to up to ~800 K at 12 $\mu m$ projectile impacts. In thermally-activated plastic flow, it is expected that a temperature rise will soften the material. Therefore, the discrepancies between FEM predictions and experimental observations likely indicate a change in material behavior under the extreme conditions imposed by LIPIT.

To further clarify the distinction between the two testing methods, we examine how the impact energy is partitioned during LIPIT loading. For that, we idealize the impact as a rigid spherical projectile of mass $m$ and initial velocity $v_i$ striking a deformable target with the density $\rho$ occupying volume $V$. The rigidity assumption is justified since projectile deformation is negligible under our impact conditions (SI Section 4). The impact initiates a sequence of energy conversions in which the projectile's initial kinetic energy is partitioned into (i) kinetic energy of the target's material points, (ii) elastic strain energy stored in the target, and (iii) energy dissipated by plastic deformation of the target. During the event, target material points evolve with velocity $\boldsymbol{v}(\mathbf{X}, t)$, stress $\boldsymbol{\sigma}(\mathbf{X}, t)$, and strain $\boldsymbol{\varepsilon}(\mathbf{X}, t)$ until time $t = t_f$, when the projectile's motion reverses due to partial elastic recovery of the target, and it rebounds with velocity $v_r$. Conservation of energy at this instant can be written as

$$\begin{aligned}\frac{m(v_i^2 - v_r^2)}{2} &= \int_V \frac{1}{2}\rho\, |\boldsymbol{v}|^2\, dV + \int_V \int_0^{t_f} \boldsymbol{\sigma} : \frac{\partial \boldsymbol{\varepsilon}}{\partial t}\, dt\, dV \\ &= \int_V \frac{1}{2}\rho\, |\boldsymbol{v}|^2\, dV + \int_V \int_0^{t_f} \boldsymbol{\sigma} : \frac{\partial \boldsymbol{\varepsilon}^{pl}}{\partial t}\, dt\, dV + \int_V \frac{1}{2}\boldsymbol{\sigma} : \boldsymbol{\varepsilon}^{el}\, dV\end{aligned} \quad 15)$$

where the total strain is decomposed into its plastic ($\boldsymbol{\varepsilon}^{pl}$) and elastic ($\boldsymbol{\varepsilon}^{el}$) components. The first term on the right represents the kinetic energy stored transiently during deformation (an inertial contribution), the second term encompasses the plastic work done during the deformation, while the third term is the elastic strain energy stored in the material due to unrecovered residual stresses and increased dislocation length.

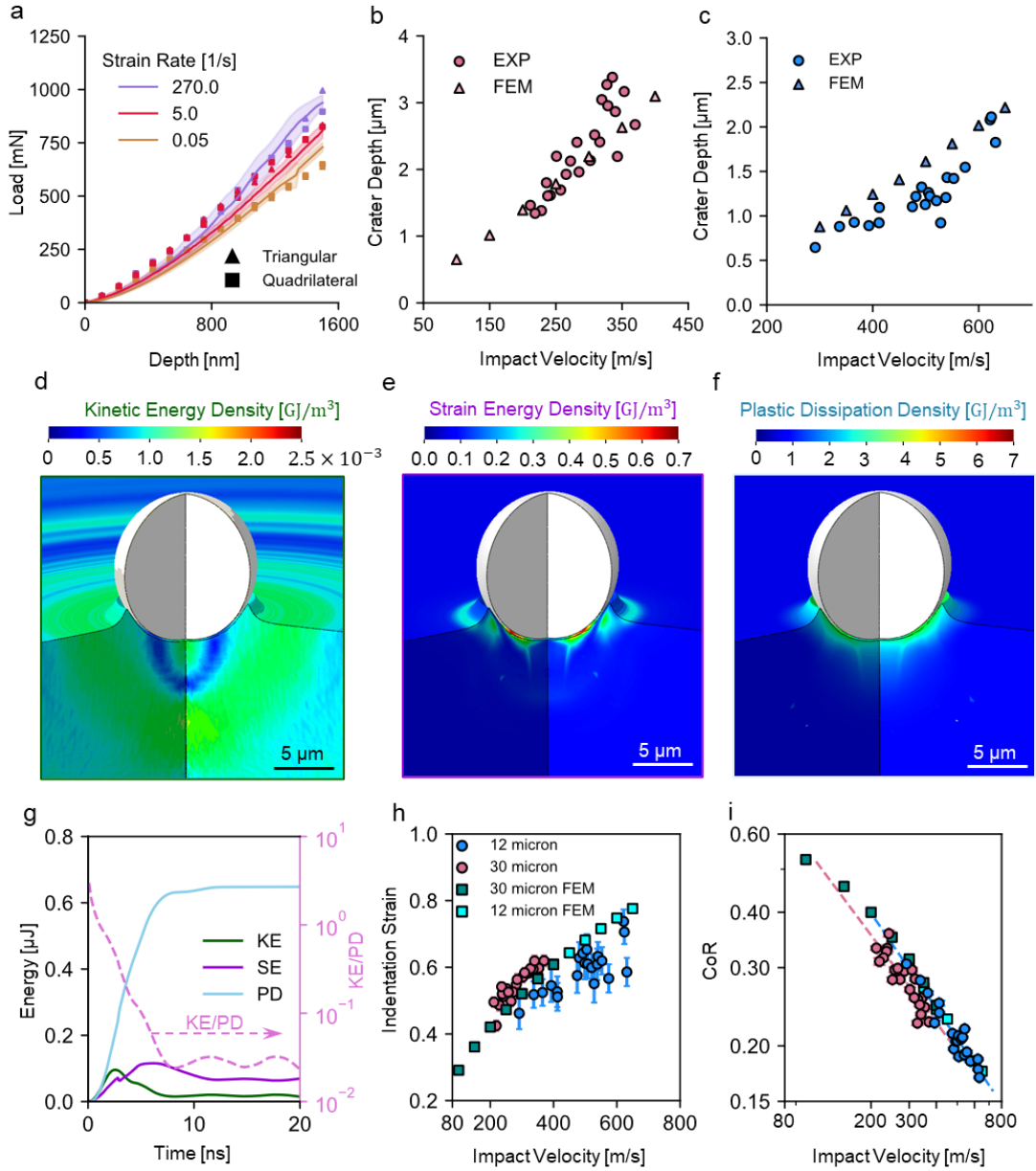

**Fig. 6. Finite Element Results.** (a) Mean load-depth curves obtained in the experiment (shadow regions show standard deviation) and calculated with FEM; Crater Depth versus velocity for (b) 30 μm and (c) 12 μm projectile impact; (d), (e), and (f) plastic dissipation (PD), kinetic energy (KE, inertial term), and strain energy (SE) densities, respectively, inside the steel target during the LIPIT impact of 12 μm projectile at 650 m/s at the time of the rebound (5-10 ns); (g) The corresponding evolution of energy terms; (h) Indentation strain and (i) Coefficient of Restitution (double logarithmic scale) developed as a function of impact velocity.

This suggests that the measured change in kinetic energy of the projectile upon impact ($\frac{m(v_i^2 - v_f^2)}{2}$) under LIPIT does not reflect the resistance of the material's plastic dissipation alone, but also contains the inertial term $\int_V \frac{1}{2} \rho |\boldsymbol{v}|^2 \, dV$ and unrecovered elastic strain energy $\int_V \frac{1}{2} \boldsymbol{\sigma} : \boldsymbol{\varepsilon}^{el} \, dV$. A large enough inertial term might propagate through an unbounded material medium and, instead of dissipating by plastic deformation alone, dissipate at the boundaries [56]. On the other hand, during nanoindentation at lower strain rates, the tip velocity is several orders of magnitude lower than during the LIPIT impact. Therefore, the SRS upturn observed in LIPIT might have a purely inertial origin, as it is suggested in Ref. [53].

To study the inertial effects, present in LIPIT but not in NI, we investigated how the projectile's kinetic energy dissipates during the supersonic LIPIT impact using our FEM model. The resultant kinetic energy ($\frac{1}{2}\rho |v|^2$), strain energy ($\frac{1}{2}\sigma : \varepsilon^{el}$), and plastic dissipation ($\int_0^{t_f} \sigma : \frac{\partial \varepsilon^{pl}}{\partial t} dt$) density fields are illustrated in Fig. 6(d-f). Unlike localized plastic dissipation and strain energy, the kinetic energy density spreads over a large volume and is several orders of magnitude smaller than plastic dissipation, due to spherical wave propagation and impact delocalization. The corresponding energy histories, obtained by integrating the density fields over the substrate volume, are plotted in Fig. 6g. Rebound occurs at ~5-10 ns after impact, coinciding with a drop in strain energy as the substrate elastically recovers. At this moment, the kinetic energy is negligible relative to the total accumulated plastic dissipation (≈2–3% of it).

Together, the analysis of the impact-energy flow and hardness estimated from the simulations indicates that the SRS upturn is unlikely to arise from inertial effects arising from differences in testing methodologies but rather is an intrinsic property of the material. Two key consequences of this SRS are the scale-dependent nominal indentation strain and the Coefficient of Restitution (CoR). Nominal indentation strain is the geometric quantity that was previously defined for NI experiments, and it is proportional to the ratio of the crater radius to the indenter radius $a$ and $R$ ($\varepsilon \propto a/R$) [57]. We employ the same definition for LIPIT impact. Using Johnson's model [58], for an elastic-perfectly plastic material—which LCS follows (Fig. 1e)—during spherical indentation, the indenter load $P$ is expressed as a function of $h$ by:

$$P = \frac{0.81}{6} \frac{\pi^3 R p_{0y}^3}{Y^2} h \qquad 16)$$

where $Y$ is the target yield strength (~1123 $MPa$) and $p_{0y} \approx 1.6Y$ is the contact pressure at the yield of the material. Integration of Eq. 16 over $h$ results in the total work done, which in this analytical model is assumed to be equal to the kinetic energy of the rigid projectile with mass $m$, density $\rho_s$ and the radius $R$ in the LIPIT impact with the initial velocity $V_i$. Hence,

$$\frac{mV_i^2}{2} = \frac{0.81}{6} \frac{\pi^3 R p_{0y}^3}{Y^2} \frac{h_f^2}{2} \qquad 17)$$

with $m = \frac{4}{3}\rho_s \pi R^3$, $h_f$ is the indentation depth. Therefore,

$$\left(\frac{h_f}{R}\right)^2 = \frac{9.876 \rho_s V_i^2 Y^2}{\pi^2 p_{0y}^3} \qquad 18)$$

This relationship shows that the ratio of the crater's depth to the impactor's radius during the impact onto a semi-infinite body remains independent of the impactor's size. Based on the impact geometry, the nominal indentation strain can be expressed as $\varepsilon = \sqrt{2\left(\frac{h_f}{R}\right) - \left(\frac{h_f}{R}\right)^2}$ (derivation is provided in SI Section 5) and therefore is also independent of the impactor's size. The analytical model is consistent with our FEM results: for both the 12 $\mu m$ and 30 $\mu m$ projectiles at impact velocities of ~300–400 $m\ s^{-1}$, the calculated indentation strain almost coincides, indicating a size-independency (Fig. 6h). However, the experimental results reveal the opposite, suggesting that, for a fixed velocity, a smaller projectile results in a lower indentation strain. The second manifestation of strain-rate sensitivity appears in the CoR ($e$) behavior. Fig. 6i shows that CoR decreases with impact velocity following a power-law scaling: $e \propto V_i^{-0.65}$ for 30 $\mu m$ and $e \propto V_i^{-0.71}$ for 12 $\mu m$ projectiles. For the impact of the elastic sphere on the elastic-perfectly plastic semi-infinite target, Wu's model [59–61] predicts a velocity-dependent but scale-invariant trend:

$$e = 1.394 \frac{Y^{3/4}}{\rho^{1/2} E^{*1/2}} V_i^{-0.5} \qquad 19)$$

where $E^*$ the reduced modulus. Similar scale-invariant behavior of CoR was also observed in finite element calculations (Fig. 6i). Yet our experimental results show that for the same $V_i$, smaller projectiles exhibit higher CoR.

These discrepancies in nominal indentation strain and CoR behavior are likely due to an SRS of the material, which results in hardening of the material, leading to the reduction of indentation strain and an increase in CoR. While logarithmic thermally-activated kinetics can qualitatively account for the rate-dependent trends in nominal indentation strain and CoR, finite-element simulations incorporating only logarithmic strain-rate hardening show no appreciable size dependence. This is likely because the material's rate sensitivity remains modest within this framework and is further offset by softening associated with the more localized deformation during the 12 $\mu m$ impact. Together, these results suggest that the pronounced size effects are likely due to the upturn in SRS at high strain rates, which conventional thermally activated flow-stress models alone cannot capture.

### 3.3. Decoupling microstructural evolution from mobility effects through re-indentation

The results demonstrate that the projectile's kinetic energy in LIPIT is mainly dissipated through rate-dependent plastic flow of the material. We now evaluate whether this plastic flow is controlled by the rate-dependent reduction in dislocation mobility at high velocities or by the material's microstructural evolution through crater re-indentation as shown in Supplementary Video S1, insets in Fig. 5c and Fig. 4a.

Fig. 5a shows the re-indention hardness in the crater regions as a function of the prior strain rate used to produce these craters in LCS material. The horizontal line at 4.12 $GPa$ indicates the average flat surface hardness without any deformation, measured with the cube-corner tip. The difference between in-situ and re-indentation hardness reflects the rate-dependent resistance to dislocation motion. Similarly, the difference between re-indentation and pristine surface hardness is a measure of the microstructural evolution contribution to the effective hardness. At lower strain rates, the in-situ and re-indentation hardness values are nearly identical, indicating that the resistance to dislocation glide is small. With increasing strain rate, the systematic widening of the gap between in-situ and re-indentation hardness is indicative of a growing contribution of rate-dependent dislocation mobility relative to structural evolution. Therefore, the reduction in dislocation mobility at higher velocities are likely reasons for the increased strain-rate sensitivity at extreme strain-rates achieved in LIPIT.

In contrast, the reindentation hardness exhibits only a weak dependence on prior applied strain-rate. As shown in Fig. 8d, the difference between the re-indentation hardness and the pristine hardness increases modestly—from ~0.5 to ~2 $GPa$—despite spanning nearly ten orders of magnitude in strain rate. The scatter in the results further reflects the influence of local microstructural heterogeneities, such as grain orientation, size, and dislocation density, present in LCS at the scale of the cube-corner indentation, which affect both the accumulation and distribution of defects during impact and, consequently, the re-indentation hardness measured afterwards.

Motivated by this observation, we investigated the relative roles of dislocation multiplication and velocity effects in 99.95% pure iron, which has a much coarser microstructure and a strong texture, resulting in more homogenous structure locally with fewer pre-existing defects than in the LCS material. Fig. 5c presents the pure iron's dynamic hardness measured during nominally 12 $\mu m$ projectile impacts, along with the corresponding re-indentation hardness. Notably, despite a quasi-static hardness of only ~1.88 $GPa$ (obtained with a cube-corner indenter), the dynamic hardness of pure iron under extreme impact conditions increases by a factor of ~4.25, reaching ~8 $GPa$—substantially more pronounced than the ~2.7× increase observed for the LCS material. As shown in Fig. 5c, the post-impact hardness increases by an average of ~100% within pure iron craters relative to the undeformed surface, compared to an increase of <50% in LCS under identical loading conditions. Meanwhile, contributions from the resistance to glide at high velocities are comparable between the two metals: ~4.44 $GPa$ in pure iron and ~5.13 $GPa$ in LCS (averaged across the strain-rate range tested).

As a result, the contribution of the reduced dislocation mobility at high velocities is responsible for the increased measured hardness at high strain rates. These results agree with the bulk-scale behavior of other BCC metals such as Ta and Ta-alloy deformed at extreme strain rate, whereby the strain-rate dependent lattice friction affects the strain-rate hardening, and subsequent reloading does not show a significant increase in flow stress [62]. Our re-indentation hardness measurements suggest that this trend remains valid even beyond the SRS upturn.

### 3.4. Microstructure-Property Relationships

To investigate the origins of the interplay between hardness, strain rate, and initial microstructure, we conducted post-mortem electron backscatter diffraction (EBSD) analysis of impact craters formed under various loading conditions summarized in Fig. 7.

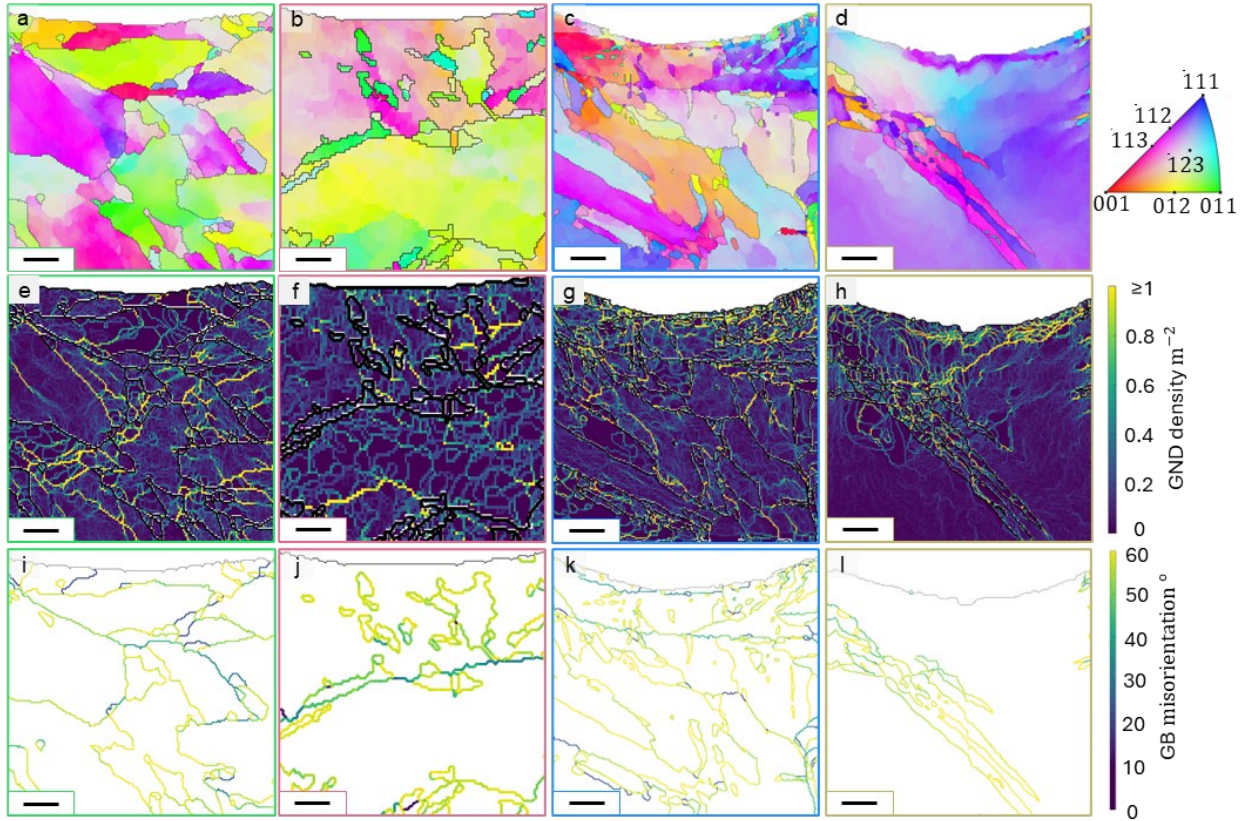

**Fig. 7. Microstructural evolution after indentation.** (a-d) EBSD orientation maps taken beneath the impact craters after the NI of the LCS at 270 $s^{-1}$ (a), LIPIT impact on LCS at ~6 × $10^6$ (30 − $\mu m$ diameter projectile at 193 $m\ s^{-1}$) (b), LIPIT impact at ~4 × $10^7\ s^{-1}$ (~12 − $\mu m$ diameter projectile at ~545 $m\ s^{-1}$) (c), and the LIPIT impact on pure iron at ~4 × $10^7\ s^{-1}$ achieved with ~12 − $\mu m$ diameter projectile at ~645 $m\ s^{-1}$ (d). (e-h) the corresponding intragranular geometrically necessary dislocation density and (i-l) intergranular misorientation maps. Scale bars correspond to 750 nm.

We first analyze the changes in microstructure due to strain rate under a constant applied deformation. For this, we fix the geometry by considering craters from the nominally ~30 $\mu m$ projectile impact in LIPIT that are topographically similar to those observed after NI, as shown in Fig. 8b-i and b-ii. Small differences between indenter tips and similar indentation depths ensure nearly identical crater geometries (Fig. 8b-ii), despite the strain-rate spanning eight orders of magnitude from ~5 × $10^{-2}$ to ~$10^6\ s^{-1}$. Fig. 7a shows the orientation map of a sample indented at ~270 $s^{-1}$ with NI while Fig. 7b shows that of a sample impacted at ~8 × $10^6\ s^{-1}$ in LIPIT. In addition, Fig. S7(a, d) shows the orientation maps of LCS deformed at 0.05 and 5 $s^{-1}$ under NI. These figures indicate that deformation proceeded primarily via dislocation motion and interactions, which resulted in crystal rotation. Based on orientation data, the Geometrically Necessary Dislocation (GND) density was estimated as described in SI Section 6. The GND density maps shown in Fig. 7(e, f), and Fig. S6(b, e) do not show significant differences in dislocation structures, with the GNDs being homogeneously distributed across the deformation zones with no significant localizations close to a crater region and hence no formation of new grains (Fig. 7(i-j) and Fig. S6(c, f)).

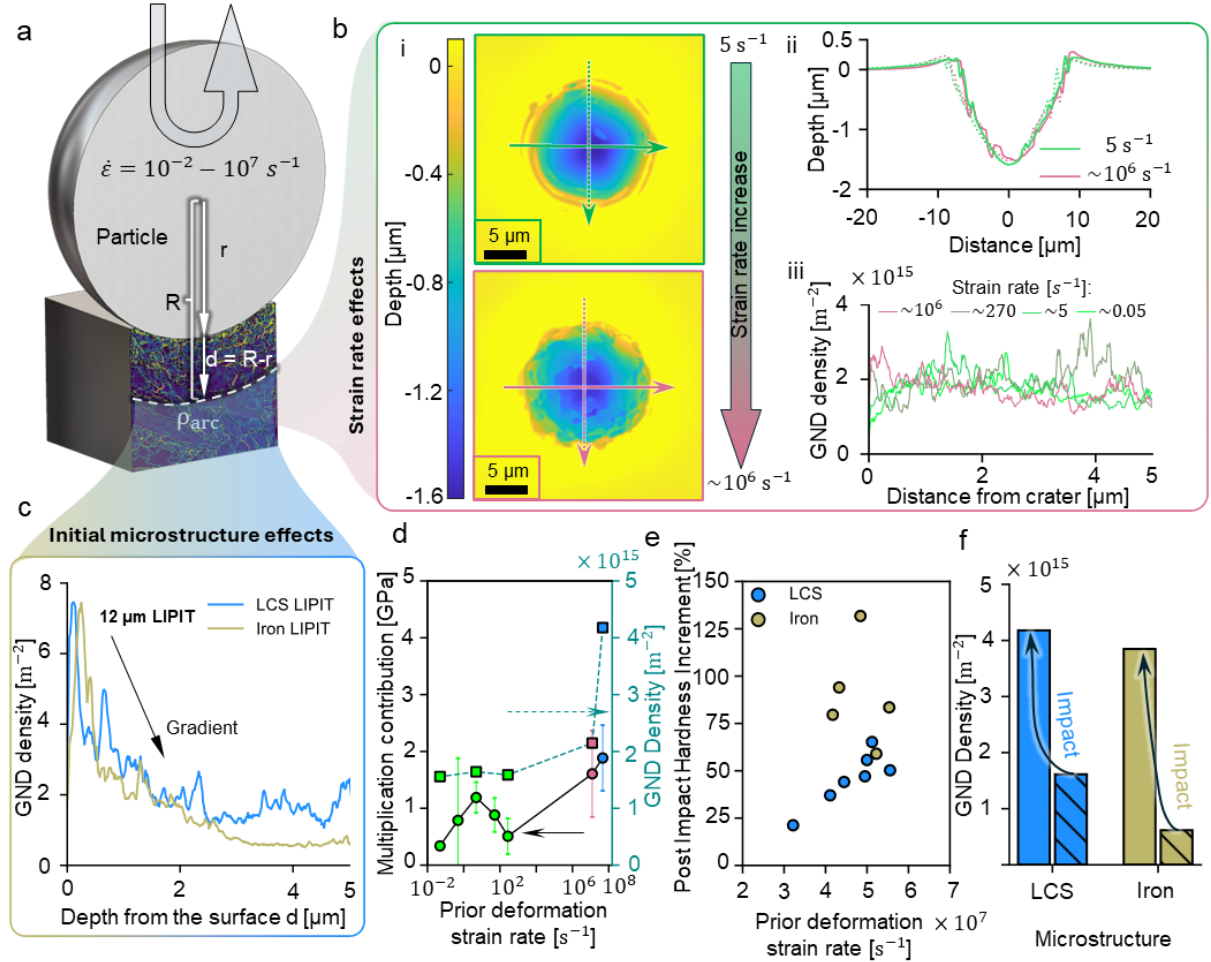

**Fig. 8. Process-structure-property relationship in the dynamic impact.** (a) Calculation of the arc-averaged GND density after the indentation; (b) strain rate effects on the microstructural evolution, where the craters with identical profiles obtained at a broad range of strain rates (b-i and b-ii) show the same dislocation density distribution after the impact (b-iii); (c) post-mortem dislocation density variation formed under 12 $\mu m$ projectile LIPIT impact at ~545 and ~645 $m\ s^{-1}$ in LCS and pure iron, suggesting the strain-induced formation of a gradient; (d) Dislocation density within 1 $\mu m$ beneath the crater, along with its contribution to hardness as a function of the prior deformation strain rate; (e) Comparison of the increase in dislocation density and relative re-indentation hardness in LCS and pure iron.

To quantify the relationship between impact-induced microstructural evolution and post-impact hardness, we calculated the average GND density along arcs of constant radius beneath the crater $\rho_{arc}$ (Fig. 8a) with the distance $d$ from the crater surface, as shown in Fig. 8b-iii. Remarkably, these microanalyses indicate that despite significant differences in strain-rate, the arc-averaged GND density remains unchanged. This correlates with the strain-rate-invariant re-indentation hardness, suggesting that its origin is likely associated with microstructural evolution weakly dependent on the strain rate.

The GND density represents only a fraction of the total dislocation density, with the remaining being statistically stored dislocations (SSDs). Collectively, GNDs exhibit a nonzero net Burgers vector within the probed volume, whereas SSDs consist of dislocation populations of opposite sign within the same probed volume that sum to zero net Burgers vector and therefore do not contribute to local rotation [63]. As we demonstrate in the SI Section 6 and as discussed in [64], decreasing the step and probe size increases the fraction of the total dislocation content that contributes to measurable lattice curvature in EBSD, such that in the infinitesimal-probe/step limit, all dislocations would be classified as GNDs. Accordingly, the measured GND density provides a lower-bar estimate of the total dislocation density.

We then analyze how the materials with different initial microstructures behave under loading conditions imposed by LIPIT. For this, we examine the impact of a 12 $\mu m$ projectile on both pure iron and LCS subjected to LIPIT, with an initial velocity of around 645 and 545 $m\ s^{-1}$, respectively, resulting in similar nominal strain rates. The orientation maps for LCS and pure iron, shown in Fig. 7(c and d), suggest that, unlike in NI, the LIPIT impact of the 30 $\mu m$ projectile, there is more significant dislocation structure development leading to crystal rotation immediately below the crater surface. This microstructure features a ~100 nm thin layer of grains elongated tangentially along the crater circumference, forming low-angle grain boundaries (Fig. 7k). Impact on pure iron—despite the crater being smaller in size than a single grain of the polycrystalline structure—results in a similar but incomplete formation of a low-angle grain boundary. The associated rotation produces an orientation in which the $\langle 1\bar{1}2 \rangle$-type lattice vector is directed normally to the lamella surface.

These reorientations are mediated by the GNDs near the crater surface (Fig. 7(g, h)) due to increased local strain at higher impact velocities, which creates a spatial gradient in GND density in both LCS and pure iron (Fig. 8c). The gradient leads to a higher GND density beneath the crater, averaging ~$4 \times 10^{15}\ m^{-2}$ in LCS, explaining a slight increment in re-indentation hardness (Fig. 8d). A similar effect was observed in pure iron deformed under identical conditions and showing the same trend in the spatial distribution of the GND density with a strong gradient near the crater surface (Fig. 8c).

The differences in material behavior, in which hardness of pure iron after deformation increases more significantly than that of the LCS, likely originate from variations in the initial GND density. While a high initial GND density (~$2 \times 10^{15}\ m^{-2}$) and fine-grained structure results in initially high hardness and mechanical properties, they also limit the material's ability to strain harden [65]. This is likely to happen as a result of dislocation annihilation caused by the interaction of oppositely signed dislocations close to each other within a high-density network of LCS [21].

Consequently, the dislocation storage driven by multiplication is partially suppressed by annihilation mechanisms in LCS. This behavior is consistent with our mechanical testing of LCS using the Kolsky bar apparatus (Fig. 1e), which showed very little strain hardening, suggesting a reduced role for dislocation multiplication during plastic flow. We demonstrated that this behavior persists over a broad range of strain rates, from quasi-static to highly dynamic, as observed in LIPIT, resulting in a smaller increment in hardness that is nearly independent of the strain rate (Fig. 8d). Conversely, a lower dislocation density of ~$6 \times 10^{14} \ m^{-2}$ and a larger grain size in pure iron may promote multiplication to a significantly greater extent than annihilation. This hypothesis aligns with the observation of a larger increase in GND density and re-indentation hardness in iron relative to the undeformed state, compared to the LCS case (Fig. 8(e, f)). Therefore, the larger hardness increases within the impact crater after re-indentation relative to the pristine surface in pure iron, compared to LCS, is likely due to a more pronounced relative increase in GND density.

## 4. Conclusions

In summary, we evaluated the contributions of microstructural evolution and dislocation mobility to the measured hardness as a function of strain rate for two BCC materials, LCS and pure iron. Through a newly developed re-indentation protocol and detailed microstructural analysis across all tested strain levels, we found that the observed SRS and its upturn in LCS are likely due to the reduced dislocation mobility at high velocities. In contrast, microstructural evolution is nearly insensitive to the applied strain-rate and instead depends more strongly on the initial microstructure. Specifically, a microstructure with an initially low dislocation density exhibits a larger increase in local dislocation content, thereby enhancing the multiplication-driven hardening component. Our results indicate that dynamic deformation processes, which improve mechanical properties, are more effective on materials having lower dislocation density, such as pure iron in this study. In contrast, materials like LCS are better suited for applications where invariant properties are desirable across broad strain-rate regimes.

**CRediT authorship contribution statement**

**Daniyar Syrlybayev:** Writing – original draft, Conceptualization, Data curation, Formal analysis, Investigation, Methodology, Validation. **Lavanya Raman:** Investigation, Data Curation. **Niraj Pramod Atale**: Investigation, Data curation. **Bhanugoban Maheswaran:** Investigation, Data curation. **Siddhartha Pathak:** Writing – review and editing, Investigation. **Curt A. Bronkhorst:** Writing – review and editing, Investigation, Funding acquisition. **Ramathasan Thevamaran:** Writing – review and editing, Supervision, Resources, Project administration, Funding acquisition, Methodology, Conceptualization.

**Declaration of Competing Interest**

The authors declare that they have no known competing financial interests or personal relationships that could have appeared to influence the work reported in this paper.

**Acknowledgments**

We acknowledge financial support from the Army Research Laboratory through the Center for Extreme Events in Structurally Evolving Materials (CEESEM) under Grant No. W911NF-23-2-0073. We also acknowledge the use of facilities and instrumentation at the UW–Madison Wisconsin Center for Nanoscale Technology, which is partially supported by the Wisconsin Materials Research Science and Engineering Center (NSF DMR-2309000) and the University of Wisconsin–Madison.

**References**

[1] B.K. Sharma, H. Basker, Space Environment and Debris: A Review of Micro-Meteoroids and Orbital Debris Impact Protection, (2025). https://doi.org/10.2139/ssrn.5377596.

[2] S.G. Love, D.E. Brownlee, N.L. King, F. Hörz, Morphology of meteoroid and debris impact craters formed in soft metal targets on the LDEF satellite, International Journal of Impact Engineering 16 (1995) 405–418. https://doi.org/10.1016/0734-743X(94)00050-7.

[3] H. Dong, J. Tang, J. Zhang, W. Huang, L. Lv, Atomistic investigation on the correlation between shot peening impact energy and grain microstructure, J Mater Sci 60 (2025) 21952–21983. https://doi.org/10.1007/s10853-025-11659-y.

[4] W. Sun, X. Chu, H. Lan, R. Huang, J. Huang, Y. Xie, J. Huang, G. Huang, Current Implementation Status of Cold Spray Technology: A Short Review, J Therm Spray Tech 31 (2022) 848–865. https://doi.org/10.1007/s11666-022-01382-4.

[5] Z. Li, J. Wang, T. Ni, W. Zhang, G. Tong, X. Chen, J. Li, R. Yan, Microstructure evolution and deformation mechanism of AZ80 alloy during die forging, Materials Science and Engineering: A 869 (2023) 144789. https://doi.org/10.1016/j.msea.2023.144789.

[6] J.D. Reid, Towards the understanding of material property influence on automotive crash structures, Thin-Walled Structures 24 (1996) 285–313. https://doi.org/10.1016/0263-8231(95)00041-0.

[7] P.J. Hazell, Armour: Materials, Theory, and Design, 2nd ed., CRC Press, Boca Raton, 2022. https://doi.org/10.1201/9781003322719.

[8] D. Hull, D.J. Bacon, eds., Introduction to dislocations, 5th ed, Butterworth-Heinemann, Oxford, 2011.

[9] W. Cai, Imperfections in Crystalline Solids, 1st ed, Cambridge University Press, West Nyack, 2016.

[10] P.M. Anderson, P.M. Anderson, J.P. Hirth, J. Lothe, Theory of dislocations, Third edition, Cambridge University Press, New York, NY, USA, 2017.

[11] M.A. Meyers, K.K. Chawla, Mechanical behavior of materials, 2. ed., 4. print. with corr, Cambridge University Press, Cambridge, 2010.

[12] G.T. (Rusty) Gray, High-Strain-Rate Deformation: Mechanical Behavior and Deformation Substructures Induced, Annu. Rev. Mater. Res. 42 (2012) 285–303. https://doi.org/10.1146/annurev-matsci-070511-155034.

[13] M.A. Meyers, Dynamic Behavior of Materials, 1st ed., Wiley, 1994. https://doi.org/10.1002/9780470172278.

[14] Y. Zhang, B.L. Hackett, J. Dong, K.Y. Xie, G.M. Pharr, Evolution of dislocation substructures in metals via high-strain-rate nanoindentation, Proc. Natl. Acad. Sci. U.S.A. 120 (2023) e2310500120. https://doi.org/10.1073/pnas.2310500120.

[15] J.H. Eggert, R.F. Smith, D.C. Swift, R.E. Rudd, D.E. Fratanduono, D.G. Braun, J.A. Hawreliak, J.M. McNaney, G.W. Collins, Ramp compression of tantalum to 330 GPa, High Pressure Research 35 (2015) 339–354. https://doi.org/10.1080/08957959.2015.1071361.

[16] R.F. Smith, J.H. Eggert, R.E. Rudd, D.C. Swift, C.A. Bolme, G.W. Collins, High strain-rate plastic flow in Al and Fe, Journal of Applied Physics 110 (2011) 123515. https://doi.org/10.1063/1.3670001.

[17] S.A. Turnage, M. Rajagopalan, K.A. Darling, P. Garg, C. Kale, B.G. Bazehhour, I. Adlakha, B.C. Hornbuckle, C.L. Williams, P. Peralta, K.N. Solanki, Anomalous mechanical behavior of nanocrystalline binary alloys under extreme conditions, Nat Commun 9 (2018) 2699. https://doi.org/10.1038/s41467-018-05027-5.

[18] S. Srinivasan, S. Sharma, S. Turnage, B.C. Hornbuckle, C. Kale, K.A. Darling, K. Solanki, Role of tantalum concentration, processing temperature, and strain-rate on the mechanical behavior of copper-tantalum alloys, Acta Materialia 208 (2021) 116706. https://doi.org/10.1016/j.actamat.2021.116706.

[19] M. Hassani, D. Veysset, K.A. Nelson, C.A. Schuh, Material hardness at strain rates beyond 106 s−1 via high velocity microparticle impact indentation, Scripta Materialia 177 (2020) 198–202. https://doi.org/10.1016/j.scriptamat.2019.10.032.

[20] Y. Song, Z. Gu, C. Huang, X. Wu, Dynamic behavior of metals under laser-induced microparticle impact, International Journal of Impact Engineering 202 (2025) 105318. https://doi.org/10.1016/j.ijimpeng.2025.105318.

[21] J. Cai, C. Griesbach, S.G. Ahnen, R. Thevamaran, Dynamic Hardness Evolution in Metals from Impact Induced Gradient Dislocation Density, Acta Materialia 249 (2023) 118807. https://doi.org/10.1016/j.actamat.2023.118807.

[22] Q. Tang, J. Li, B.C. Hornbuckle, A. Giri, K. Darling, M. Hassani, Suppressed ballistic transport of dislocations at strain rates up to 109 s–1 in a stable nanocrystalline alloy, Commun Mater 6 (2025) 43. https://doi.org/10.1038/s43246-025-00757-8.

[23] R.F. Smith, R.W. Minich, R.E. Rudd, J.H. Eggert, C.A. Bolme, S.L. Brygoo, A.M. Jones, G.W. Collins, Orientation and rate dependence in high strain-rate compression of single-crystal silicon, Phys. Rev. B 86 (2012) 245204. https://doi.org/10.1103/PhysRevB.86.245204.

[24] I. Dowding, C.A. Schuh, Metals strengthen with increasing temperature at extreme strain rates, Nature 630 (2024) 91–95. https://doi.org/10.1038/s41586-024-07420-1.

[25] E.B. Zaretsky, G.I. Kanel, Yield stress, polymorphic transformation, and spall fracture of shock-loaded iron in various structural states and at various temperatures, Journal of Applied Physics 117 (2015) 195901. https://doi.org/10.1063/1.4921356.

[26] M.B. Prime, S.J. Fensin, D.R. Jones, J.W. Dyer, D.T. Martinez, Multiscale Richtmyer-Meshkov instability experiments to isolate the strain rate dependence of strength, Phys. Rev. E 109 (2024) 015002. https://doi.org/10.1103/PhysRevE.109.015002.

[27] W. Cai, V.V. Bulatov, J. Chang, J. Li, S. Yip, Dislocation Core Effects on Mobility, in: Dislocations in Solids, Elsevier, 2004: pp. 1–80. https://doi.org/10.1016/S1572-4859(05)80003-8.

[28] J. Chang, W. Cai, V.V. Bulatov, S. Yip, Dislocation motion in BCC metals by molecular dynamics, Materials Science and Engineering: A 309–310 (2001) 160–163. https://doi.org/10.1016/S0921-5093(00)01673-7.

[29] J. Chang, W. Cai, V.V. Bulatov, S. Yip, Molecular dynamics simulations of motion of edge and screw dislocations in a metal, Computational Materials Science 23 (2002) 111–115. https://doi.org/10.1016/S0927-0256(01)00221-X.

[30] G.I. Kanel, S.V. Razorenov, G.V. Garkushin, S.I. Ashitkov, P.S. Komarov, M.B. Agranat, Deformation resistance and fracture of iron over a wide strain rate range, Phys. Solid State 56 (2014) 1569–1573. https://doi.org/10.1134/S1063783414080113.

[31] A.Yu. Kuksin, A.V. Yanilkin, Atomistic simulation of the motion of dislocations in metals under phonon drag conditions, Phys. Solid State 55 (2013) 1010–1019. https://doi.org/10.1134/S1063783413050193.

[32] J. Marian, W. Cai, V.V. Bulatov, Dynamic transitions from smooth to rough to twinning in dislocation motion, Nature Mater 3 (2004) 158–163. https://doi.org/10.1038/nmat1072.

[33] C.L. Williams, C.Q. Chen, K.T. Ramesh, D.P. Dandekar, On the shock stress, substructure evolution, and spall response of commercially pure 1100-O aluminum, Materials Science and Engineering: A 618 (2014) 596–604. https://doi.org/10.1016/j.msea.2014.09.030.

[34] C.L. Williams, C.Q. Chen, K.T. Ramesh, D.P. Dandekar, The effects of cold rolling on the microstructural and spall response of 1100 aluminum, Journal of Applied Physics 114 (2013) 093502. https://doi.org/10.1063/1.4817844.

[35] R. Thevamaran, O. Lawal, S. Yazdi, S.-J. Jeon, J.-H. Lee, E.L. Thomas, Dynamic creation and evolution of gradient nanostructure in single-crystal metallic microcubes, Science 354 (2016) 312–316. https://doi.org/10.1126/science.aag1768.

[36] C. Griesbach, J. Cai, S.-J. Jeon, R. Thevamaran, Synergistic strength and toughness through impact-induced nanostructural evolutions in metals, Extreme Mechanics Letters 62 (2023) 102037. https://doi.org/10.1016/j.eml.2023.102037.

[37] C. Griesbach, J. Cai, S.-J. Jeon, R. Thevamaran, Orientation-dependent plasticity mechanisms control synergistic property improvement in dynamically deformed metals, International Journal of Plasticity 166 (2023) 103657. https://doi.org/10.1016/j.ijplas.2023.103657.

[38] A.A. Tiamiyu, E.L. Pang, X. Chen, J.M. LeBeau, K.A. Nelson, C.A. Schuh, Nanotwinning-assisted dynamic recrystallization at high strains and strain rates, Nat. Mater. 21 (2022) 786–794. https://doi.org/10.1038/s41563-022-01250-0.

[39] C. Griesbach, C.A. Bronkhorst, R. Thevamaran, Crystal plasticity simulations reveal cooperative plasticity mechanisms leading to enhanced strength and toughness in gradient nanostructured metals, Acta Materialia 270 (2024) 119835. https://doi.org/10.1016/j.actamat.2024.119835.

[40] L. Lea, L. Brown, A. Jardine, Time limited self-organised criticality in the high rate deformation of face centred cubic metals, Commun Mater 1 (2020) 93. https://doi.org/10.1038/s43246-020-00090-2.

[41] A. Rohatgi, K.S. Vecchio, G.T. Gray, Iii, A metallographic and quantitative analysis of the influence of stacking fault energy on shock-hardening in Cu and Cu–Al alloys, Acta Materialia 49 (2001) 427–438. https://doi.org/10.1016/S1359-6454(00)00335-9.

[42] H. Fan, Q. Wang, J.A. El-Awady, D. Raabe, M. Zaiser, Strain rate dependency of dislocation plasticity, Nat Commun 12 (2021) 1845. https://doi.org/10.1038/s41467-021-21939-1.

[43] D. Raabe, C.C. Tasan, E.A. Olivetti, Strategies for improving the sustainability of structural metals, Nature 575 (2019) 64–74. https://doi.org/10.1038/s41586-019-1702-5.

[44] Dassault Systèmes, ABAQUS Analysis User's Manual, n.d.

[45] H. Klippel, M. Gerstgrasser, D. Smolenicki, E. Cadoni, H. Roelofs, K. Wegener, Johnson Cook Flow Stress Parameter for Free Cutting Steel 50SiB8, (2020). https://doi.org/10.48550/ARXIV.2007.14087.

[46] J. Cai, C. Griesbach, S.G. Ahnen, R. Thevamaran, Dynamic Hardness Evolution in Metals from Impact Induced Gradient Dislocation Density, Acta Materialia 249 (2023) 118807. https://doi.org/10.1016/j.actamat.2023.118807.

[47] W.C. Oliver, G.M. Pharr, An improved technique for determining hardness and elastic modulus using load and displacement sensing indentation experiments, J. Mater. Res. 7 (1992) 1564–1583. https://doi.org/10.1557/JMR.1992.1564.

[48] D.S. Stone, K.B. Yoder, W.D. Sproul, Hardness and elastic modulus of TiN based on continuous indentation technique and new correlation, Journal of Vacuum Science & Technology A: Vacuum, Surfaces, and Films 9 (1991) 2543–2547. https://doi.org/10.1116/1.577270.

[49] J.E. Jakes, C.R. Frihart, J.F. Beecher, R.J. Moon, D.S. Stone, Experimental method to account for structural compliance in nanoindentation measurements, J. Mater. Res. 23 (2008) 1113–1127. https://doi.org/10.1557/jmr.2008.0131.

[50] D.L. Joslin, W.C. Oliver, A new method for analyzing data from continuous depth-sensing microindentation tests, J. Mater. Res. 5 (1990) 123–126. https://doi.org/10.1557/JMR.1990.0123.

[51] Y. Liu, J. Du, K. Zhang, K. Gao, H. Xue, X. Fang, K. Song, F. Liu, Orientation-Dependent Mechanical Behaviors of BCC-Fe in Light of the Thermo-Kinetic Synergy of Plastic Deformation, Materials 17 (2024) 2395. https://doi.org/10.3390/ma17102395.

[52] X. Zhou, S. He, J. Marian, Cross-kinks control screw dislocation strength in equiatomic bcc refractory alloys, Acta Materialia 211 (2021) 116875. https://doi.org/10.1016/j.actamat.2021.116875.

[53] Z. Ghasemi, T. Dos Santos, J.A. Rodríguez-Martínez, A. Srivastava, Inertial effect on dynamic hardness and apparent strain-rate sensitivity of ductile materials, Journal of

the Mechanics and Physics of Solids 180 (2023) 105418. https://doi.org/10.1016/j.jmps.2023.105418.

[54] T. Dos Santos, A. Srivastava, J.A. Rodríguez-Martínez, The combined effect of size, inertia and porosity on the indentation response of ductile materials, Mechanics of Materials 153 (2021) 103674. https://doi.org/10.1016/j.mechmat.2020.103674.

[55] Z. Rosenberg, R. Kositski, Y. Ashuach, V. Leus, A. Malka-Markovitz, On the upturn phenomenon in the strength vs. strain-rate relations of metals, International Journal of Solids and Structures 176–177 (2019) 185–190. https://doi.org/10.1016/j.ijsolstr.2019.06.015.

[56] Y. Dharmadasa, N. Jaegersberg, A. Kim, J. Cai, R. Thevamaran, Momentum-Transfer Framework Unifies High-Velocity Impact and Failure Across Materials, Geometries, and Scales, (2025). https://doi.org/10.48550/ARXIV.2510.26360.

[57] S. Pathak, S.R. Kalidindi, Spherical nanoindentation stress–strain curves, Materials Science and Engineering: R: Reports 91 (2015) 1–36. https://doi.org/10.1016/j.mser.2015.02.001.

[58] K.L. Johnson, Contact Mechanics, 1st ed., Cambridge University Press, 1985. https://doi.org/10.1017/CBO9781139171731.

[59] C.-Y. Wu, C. Thornton, L.-Y. Li, Coefficients of restitution for elastoplastic oblique impacts, Advanced Powder Technology 14 (2003) 435–448. https://doi.org/10.1163/156855203769710663.

[60] C. Wu, L. Li, C. Thornton, Rebound behaviour of spheres for plastic impacts, International Journal of Impact Engineering 28 (2003) 929–946. https://doi.org/10.1016/S0734-743X(03)00014-9.

[61] C.-Y. Wu, L.-Y. Li, C. Thornton, Energy dissipation during normal impact of elastic and elastic–plastic spheres, International Journal of Impact Engineering 32 (2005) 593–604. https://doi.org/10.1016/j.ijimpeng.2005.08.007.

[62] G.T. Gray, K.S. Vecchio, Influence of peak pressure and temperature on the structure/property response of shock- loaded Ta and Ta-10W, Metall Mater Trans A 26 (1995) 2555–2563. https://doi.org/10.1007/BF02669413.

[63] O. Muránsky, L. Balogh, M. Tran, C.J. Hamelin, J.-S. Park, M.R. Daymond, On the measurement of dislocations and dislocation substructures using EBSD and HRSD techniques, Acta Materialia 175 (2019) 297–313. https://doi.org/10.1016/j.actamat.2019.05.036.

[64] A.C. Leff, C.R. Weinberger, M.L. Taheri, Estimation of dislocation density from precession electron diffraction data using the Nye tensor, Ultramicroscopy 153 (2015) 9–21. https://doi.org/10.1016/j.ultramic.2015.02.002.

[65] R.B. Figueiredo, T.G. Langdon, Deformation mechanisms in ultrafine-grained metals with an emphasis on the Hall–Petch relationship and strain rate sensitivity, Journal of Materials Research and Technology 14 (2021) 137–159. https://doi.org/10.1016/j.jmrt.2021.06.016.

# Supporting Information for

# Decoupling dislocation multiplication and velocity effects in metals at extreme strain rates

Daniyar Syrlybayev[1], Lavanya Raman[1], Niraj Pramod Atale[2], Bhanugoban Maheswaran[1], Siddhartha Pathak[2], Curt A. Bronkhorst[1], Ramathasan Thevamaran[1,*]

*Department of Mechanical Engineering, University of Wisconsin-Madison, Madison, WI 53706*

[2]*Department of Materials Science and Engineering, Iowa State University, Ames, IA 50011, USA*

[*] Corresponding author: thevamaran@wisc.edu (RT)

1. **Kolsky Bar Testing**

To characterize the moderate-strain-rate constitutive response of quenched and tempered martensitic low carbon steel (LCS), we employed a custom-built Kolsky (Split-Hopkinson bar) experimental set-up. Here, we tested cylindrical LCS samples of two different orientations—normal direction (ND) and transverse direction (TD)—with a diameter of 3.2 mm and a thickness of 1.9 mm.

The Kolsky bar setup utilized high-strength maraging steel bars (MAR 350) with a diameter of 13 mm, procured from the Steel Service Aerospace Corp. The incident and transmitter bars were 1500 mm and 1100 mm in length, respectively, while the striker bar was 300 mm in length. To capture the signals of the stress-wave propagation through the bar with high fidelity, we utilized resistive strain gauges (Micro-Measurements, CEA-06-250UNA-350). To measure the incident wave signal and reflected wave signal, we mounted the strain gauges on the incident bar at a position 750 mm from the specimen-incident bar interface. To measure the transmitted wave signal, we mounted the strain gauges on the transmitter bar 550 mm from the sample-transmitter bar interface.

We accelerated the striker bar using a pressurized gas-driven projectile launch system powered by an air compressor, enabling precise control of striker velocities. By tuning the striker's launch pressure, thereby the striker velocity, we achieved compressive strain rates ranging from approximately 2,000 s⁻¹ to 3,000 s⁻¹. We used aluminum discs of 2 mm thickness and 5 mm diameter as pulse shapers to achieve near-constant strain rate and dynamic stress equilibrium during loading.

2. **Calculation of velocity from high-speed imaging**

We developed an algorithm for processing the stroboscopic images obtained during LIPIT impact with the high-speed camera system (Fig. S1a) to determine the displacement $x(t)$ and the velocity functions $v(t)$. For that we consider the flight of the spherical projectile with a diameter $d_p$ while it is reaching its terminal velocity in the air with the room-temperature density $\rho_{air} = 1.2 \ kg/m^3$ and viscosity $\mu_{air} = 1.8 \times 10^{-5} \ Pa*s$ [1]. Neglecting gravity effects (which are few orders of magnitude smaller than air drag making particle effectively airborne), the ordinary differential equation (ODE) describing its motion is given by:

$$m\frac{dv}{dt} = \frac{\rho_s \pi d_p^3}{6}\frac{dv}{dt} = -\frac{1}{2} * \rho_{air} v^2 C_d * \frac{\pi D^2}{4} \qquad (1)$$

$C_d$ is the velocity-dependent air drag coefficient. The empirical relationship between the air drag coefficient $C_d$ with the Reynolds number $Re = \rho_{air} v D / \mu_{air}$ shown in Fig. S1a [2]. Considering the range of test velocities and sizes of projectiles, we approximate the relationship between the Reynolds number in the range 10-1000 and the $C_d$ with the general power law $C_d = A \times Re^n$ (Fig. S1b). Substituting into Eq. 1 results

$$\frac{dv}{dt} = -Bv^{2+n} \tag{2}$$

$$\frac{d^2x}{dt^2} = -B\left(\frac{dx}{dt}\right)^{2+n} \tag{3}$$

Solving under ICs: $\frac{dx}{dt}|_{t=0} = v_0$ (initial velocity) and $x(t=0) = 0$ (initial location on the 1D reference point)

$$x(t) = \frac{(n+1)\left(Bt + \frac{1}{v_0^{n+1}(n+1)}\right)}{Bn\left((n+1)\left(Bt + \frac{1}{v_0^{n+1}(n+1)}\right)\right)^{\frac{1}{n+1}}} - \frac{1}{Bv_0^n n} \tag{4}$$

Here $B$ is the constant term and is given by

$$B = \frac{A}{2m}(\rho^{n+1}\left(\frac{d_p}{\mu_{air}}\right)^n)(\frac{\pi(d_p)^2}{4}) \tag{5}$$

As shown, the fitting values are $A = 9.66$ and $n = 0.455$. Particle diameter was measured during each test, and the value $B$ was calculated. From the obtained image (Fig. S1c) we then find the position $x(t)$ of the center of each projectile with respect to the frame X'-Y' located at the center of the first projectile. Each projectile is captured every ~256 ns corresponding to a set repetition rate of the camera. That position-time information is used to fit Eq. 4 to find the initial velocity $v_0$ (Fig. S1d). The derivative of the position function $x(t)$ corresponds to the particle velocity $v(t)$. We extrapolate it to determine the velocity of the projectile the moment projectile hits the substrate. An identical procedure is used for determining the rebound velocity (Fig. S1e).

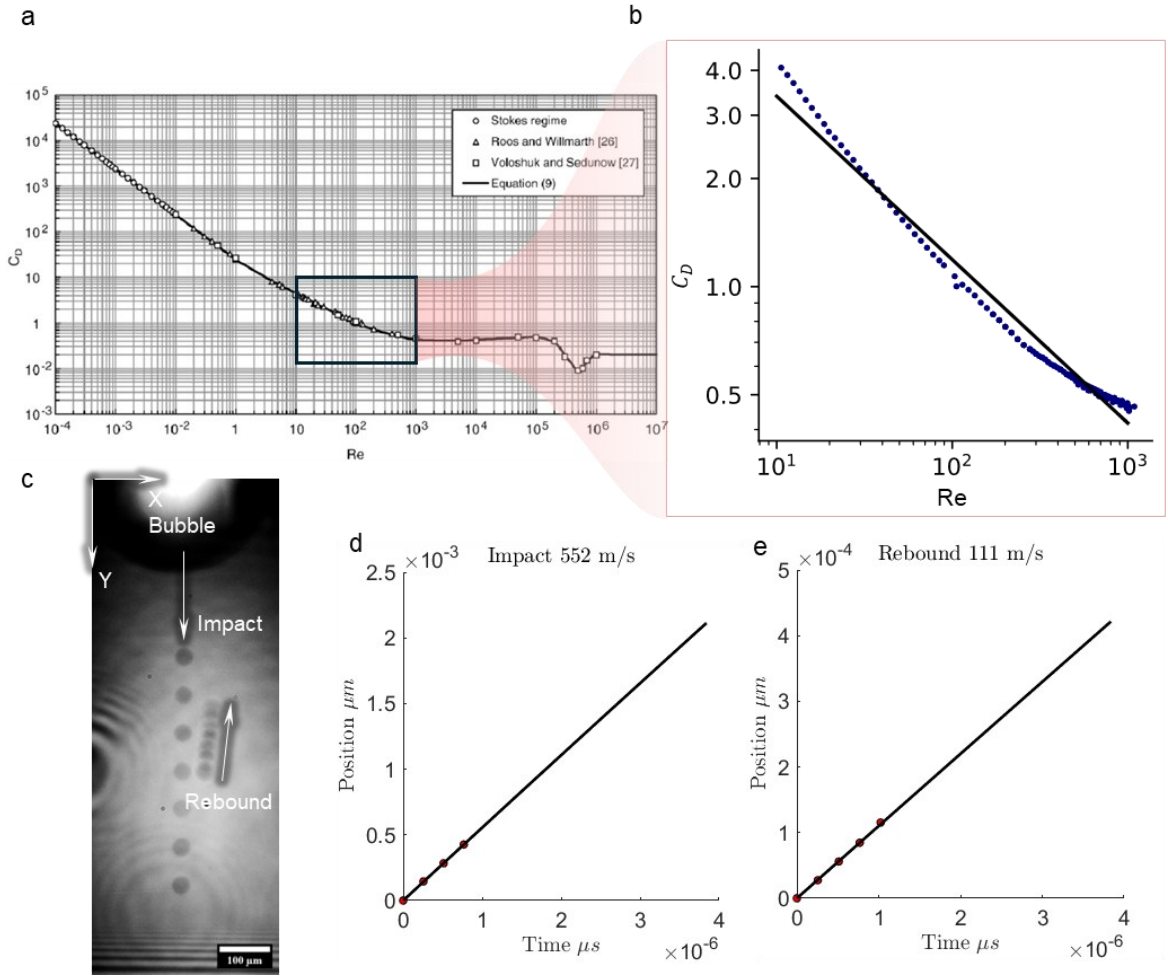

**Fig. S1. Air Drag Correction Details.** (a) Drag coefficients for a spherical projectile in air were obtained from empirical charts as a function of Reynolds number [2]. (b) Values corresponding to the Reynolds-number range of 10–1000—appropriate for our projectile size and peak velocity—were extracted and used in the particle position–time model (Eq. 4). (c) Projectile positions were obtained at discrete time intervals from high-speed camera frames. These positions were then fit using the drag-corrected model to determine the (d) impact and (e) rebound velocities at the sample surface.

### 3. Lamella Preparation and Dictionary Indexing of EBSD Data

We prepared samples for post-mortem EBSD analysis using Focused Ion Beam (FIB), as shown in Fig. S2(a-c). This process involves depositing a protective Pt layer on the crater (Fig. S2b), trench milling (Fig. S2c), lift-out, attaching the lamella to the grid, and thinning to achieve the electron channeling grain contrast (Fig. S2f). Next, the FIB-grid with lamella was attached to the EBSD holder with a 70-degree pre-tilt (Fig. S2(d, e)) for EBSD analysis. During EBSD, Kikuchi patterns were collected pre-indexed using commercial Hough Indexing from Aztec software. These patterns are stored in *.h5oina* format as a 4D array with 2D images of each pattern in each of the 2D spatial locations. Patterns were collected at 20 keV beam energy and 60 $\mu m$ aperture that allowed reducing the electron interaction volume, thereby improving the spatial accuracy of the orientations by enabling a finer probe.

Severe plastic deformation limits the performance of default Hough indexing. An example of the Hough Indexed map from 30 $\mu m$ nominal diameter projectile impact in LIPIT is shown in Fig. S2g. Therefore, we used custom dictionary indexing. For this purpose, the Python-based Kikuchipy library [3] was used.

Dictionary indexing relies on simulating the master pattern and projecting it onto the calibrated detector screen [4], as shown in Fig. S2d. In this work, we used a pre-simulated ferrite ($\alpha$-Fe, BCC) phase with an energy of 20 keV available in Kikuchipy. We then selected the 25 highest-quality patterns available across the map obtained during the EBSD scan to calibrate the pattern center. Geometrically, the pattern center corresponds to the point on the detector screen where the diffraction-induced Kikuchi sphere is tangent to the plane of the screen (Fig. S2d). Kikuchipy was used for pattern center calibration, where we employed Hough Indexing to optimize the indexing rate by simultaneously searching the crystal orientation angles and pattern center coordinates through the Nelder-Mead optimization. The second step of calibration involved dictionary indexing, where both orientation and pattern center obtained in the first step were varied locally within the neighborhood. While adjusting the orientation and pattern center, the optimization compared simulated patterns $A_{kij}$ with those obtained during the scan $B_{kij}$ to find the refined orientation and pattern center such that the Normalized Cross-Correlation (NCC) given by

$$NCC_k = \frac{(A_{kij} - \bar{A}_k)(B_{kij} - \bar{B}_k)}{\sqrt{(A_{kij} - \bar{A}_k)(A_{kij} - \bar{A}_k)}\sqrt{(B_{kpq} - \bar{B}_k)(A_{kpq} - \bar{B}_k)}} \quad (6)$$

is maximized. Here the first index $k$ in the three-dimensional array represents the pattern number, while the other indices $ij$ are pixel intensities. $\bar{A}_k$ and $\bar{B}_k$ denote the mean intensities of the pattern $k$, which are subtracted to mean-center the simulated and experimental patterns. Optimization of NCC allowed us to find the pattern center that will be used to project the simulated master pattern onto create a dictionary of patterns.

With the pattern center calibrated, we built a dictionary of patterns. To do this, we first defined the orientations of the crystals by sampling the angles defined in quaternions every 2°. Owing to the cubic symmetry, the space can be reduced to the fundamental asymmetric space, with all other orientations outside of this space having crystallographically equivalent orientations within the fundamental space. We project the master pattern for each of the orientations sampled from the fundamental space to get the dictionary of patterns. That dictionary was used to find the orientation of each experimental pattern that maximizes the NCC score.

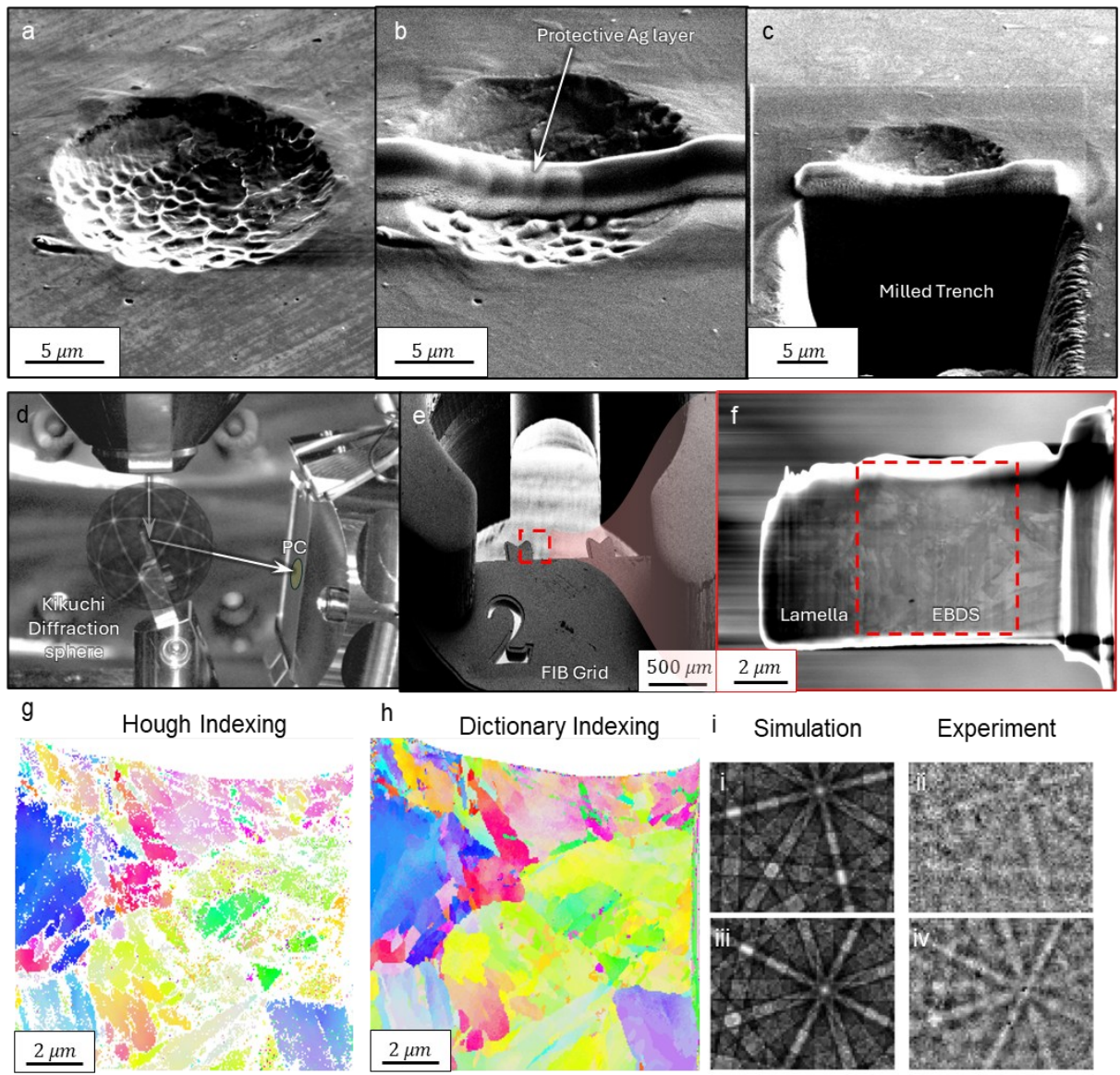

**Fig. S2 Sample Preparation and Dictionary Indexing Workflow.** (a) A 30-μm impact crater was capped with a protective Pt layer before FIB milling. (b–c) Site-specific lift-out and lamella thinning were performed, followed by final polishing. (d) EBSD measurement setup with the lamella mounted on a FIB grid (e–f) and secured on the holder. At 20 keV, the incident electron beam generates Kikuchi diffraction from the sample, producing a Kikuchi sphere whose projection is recorded by the detector. (g) Indexing performance using commercial Hough-based analysis, and (h) results from Dictionary Indexing, with (i) representative comparisons between simulated and experimental patterns.

After that procedure, the resolution of orientation is around 2° and is approximately equal to the sampling step size. To improve the indexing even further, each orientation was further optimized within 2° neighborhood using the *refine_orientation_projection_center* command that simultaneously varies local orientation and pattern center to optimize NCC further, as explained in [3].

As a result, the reindexed map is shown in Fig. S2h, along with representative Kikuchi patterns from the simulation and experiment, as shown in Fig. S2i, illustrating a close match between them.

4. **Justification of the projectile rigidity assumption**

For a rigid spherical projectile, the crater volume $V_c$ created by indentation to a depth $h$ with projectile radius $R$ is given by [5]

$$V_c = \frac{\pi h^2}{3}(3R - h) \tag{7}$$

Upon the rebound the volume of the crater would change slightly as the result of strain energy recovery in the target upon the unloading. Therefore, if the projectile can be considered rigid, the measured crater volume is expected to closely match the geometrically computed volume for a given depth. As shown in Fig. S3, the measured and computed crater volumes are nearly identical for both projectile sizes (30 $\mu m$ and 12 $\mu m$ diameter), supporting the assumption of a rigid projectile under these impact conditions.

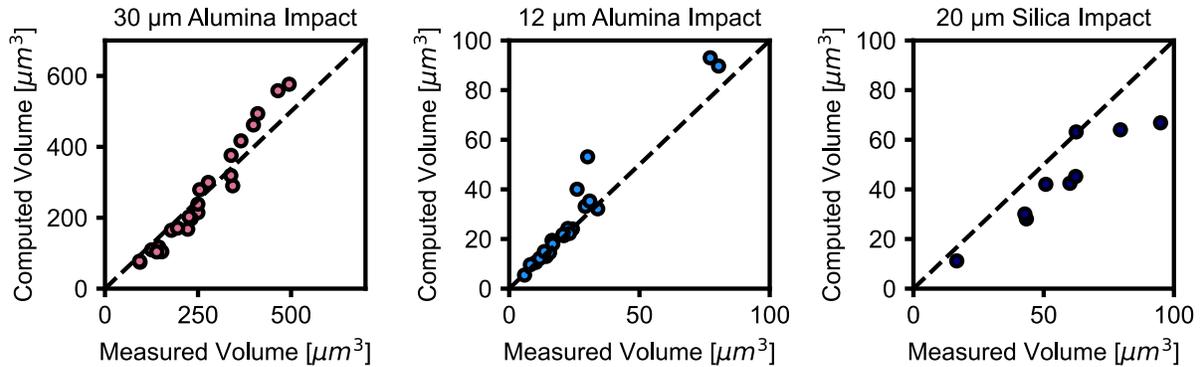

**Fig. S3.** Computed versus measured crater volume under impacts of alumina and silica

This result contrasts with our earlier silica projectile impact experiments (Fig. S3), where a significant deviation between measured and computed crater volumes was observed. In that case, the measured crater volumes are larger than the corresponding geometric predictions for a given depth, suggesting the larger crater curvature radius according to Eq. 7. This is attributed to projectile flattening and the associated increase in contact radius due to substantial projectile deformation.

5. **Estimation of the Indentation Strain**

The nominal strain in nanoindentation experiments is defined as $\varepsilon_{nom} \propto \frac{a}{R}$ where $a$ and $R$ are radii of crater and the projectile. From Fig. S4 we can express

$$a^2 = R^2 - (R - h_f)^2 = 2Rh_f - h_f^2 \tag{8}$$

Dividing both sides of equation by $R^2$ results

$$\frac{a}{R} = \sqrt{2\left(\frac{h_f}{R}\right) - \left(\frac{h_f}{R}\right)^2} \tag{9}$$

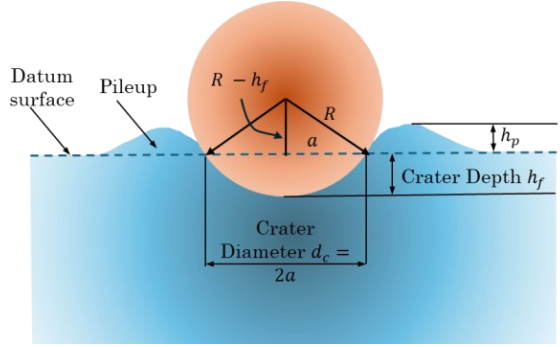

**Fig. S4.** The crater geometry during the indentation

## 6. Estimation of Geometrically Necessary Dislocation Density

We estimated the Geometrically Necessary Dislocation (GND) density using the Nye's dislocation density tensor $\alpha_{ij}$ given by [6–8]:

$$\alpha_{ij} = \sum_t b_i^t \hat{\zeta}_j^t \delta(\boldsymbol{x} - \boldsymbol{x^t}) = -\epsilon_{jlk}\frac{\partial \beta_{ik}}{\partial x_l} \tag{10}$$

where $b_i^t$ and $\hat{\zeta}_j^t$ are Burgers vector and dislocation line vector, respectively, $\delta(\boldsymbol{x} - \boldsymbol{x^t})$ is the Dirac delta function. Equation suggests that Nye's dislocation density tensor is equivalent to the curl of the elastic deformation gradient $\beta_{ik}$ which can be decomposed to its skew symmetric rotation $\omega_{ik}$ and elastic strain $\varepsilon_{ik}^{el}$ components as

$$\beta_{ik} = \varepsilon_{ik}^{el} + \omega_{ik} \tag{11}$$

In conventional EBSD the strain term $\varepsilon_{ik}^{el}$ is inaccessible, although it is possible to measure it using cross-correlation EBSD as described in [9,10]. In this work we neglect strain term and focus the rotation tensor $\omega_{ik}$ which can be expressed by:

$$\omega_{ik} = \begin{bmatrix} 0 & -\theta_3 & -\theta_2 \\ -\theta_3 & 0 & -\theta_1 \\ -\theta_2 & -\theta_1 & 0 \end{bmatrix} = \epsilon_{ikj}\theta_j \tag{12}$$

where infinitesimal rotation angle $\theta_j$ with respect to the reference are directly obtained from EBSD. In this work we used MTEX [11] to compute the Nye's dislocation density tensor from our EBSD. Following [12], we approximate GND density from the Euclidean norm by:

$$\rho_{GND} = \frac{\boldsymbol{\alpha}:\boldsymbol{\alpha}}{|\boldsymbol{b}|_2} = \frac{||\boldsymbol{\alpha}||_2}{|\boldsymbol{b}|_2} \tag{13}$$

where $|\boldsymbol{b}|_2$=0.248 nm is the magnitude of the Burgers vector.

GND density measured by EBSD reflects the local net Burgers vector content but does not capture information about dislocation dipoles whose net Burgers vector is zero, which are stored as Statistically Stored Dislocations (SSDs). As discussed in [13,14], the GND density measured by EBSD depends on the probed volume and the step size; in the limit of an infinitesimal probe volume and step size, all dislocations would be classified as GNDs [15].

We illustrate these observations experimentally by showing that decreasing EBSD step size increases the measured GND density. As shown in Fig. S5 the reduction in step size from 200 to 50 nm (Fig. S5(a-c)) reveals more intricate dislocation structure within each grain (Fig. S5(d-f)), increasing the average GND density calculated from full maps (Fig. S5g).

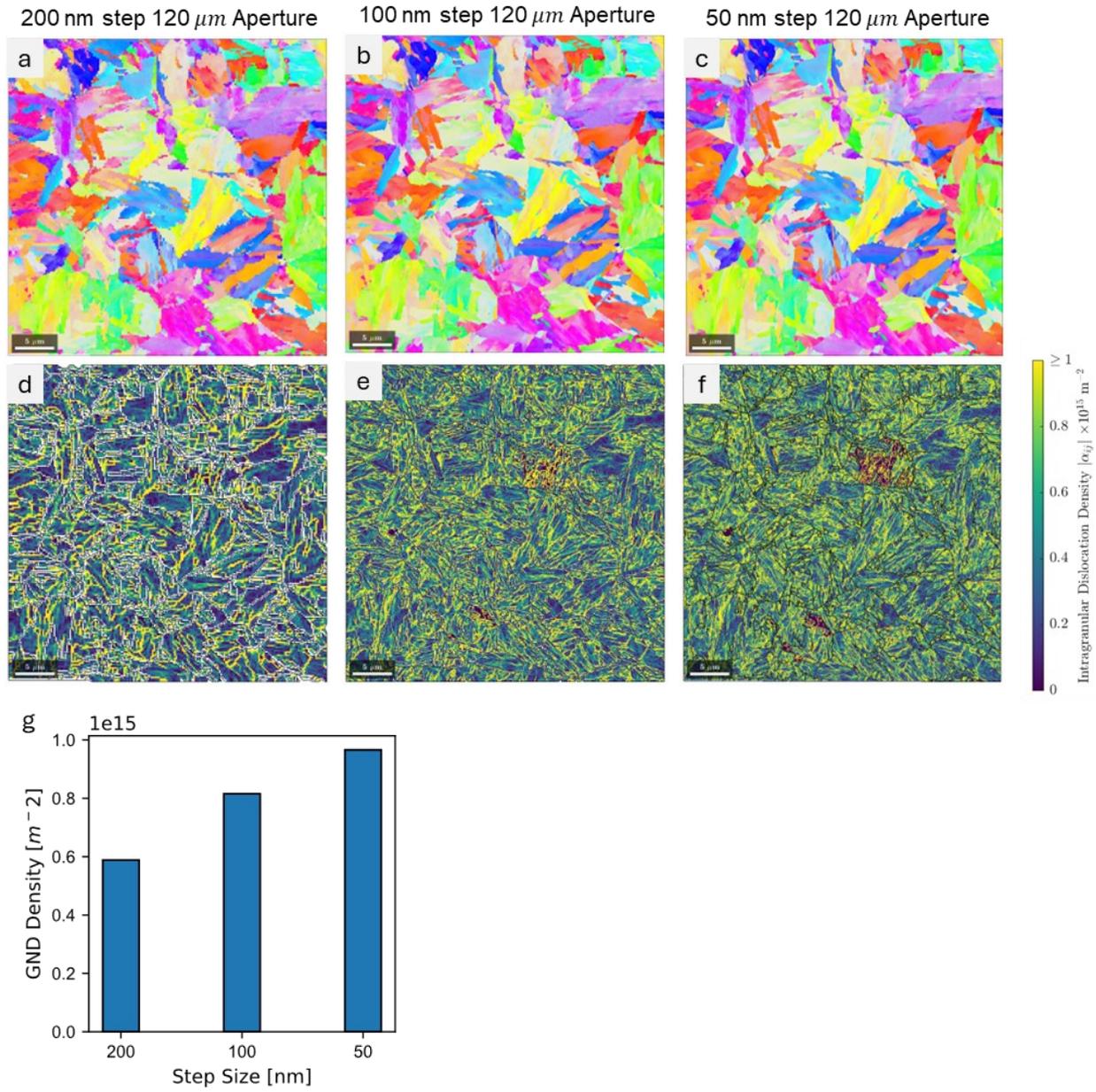

**Fig. S5. Illustration of sensitivity of the GND density to the step size.** (a-c) Orientation maps acquired with different step sizes; (d-f) corresponding dislocation density maps; (g) Average GND density calculated from maps

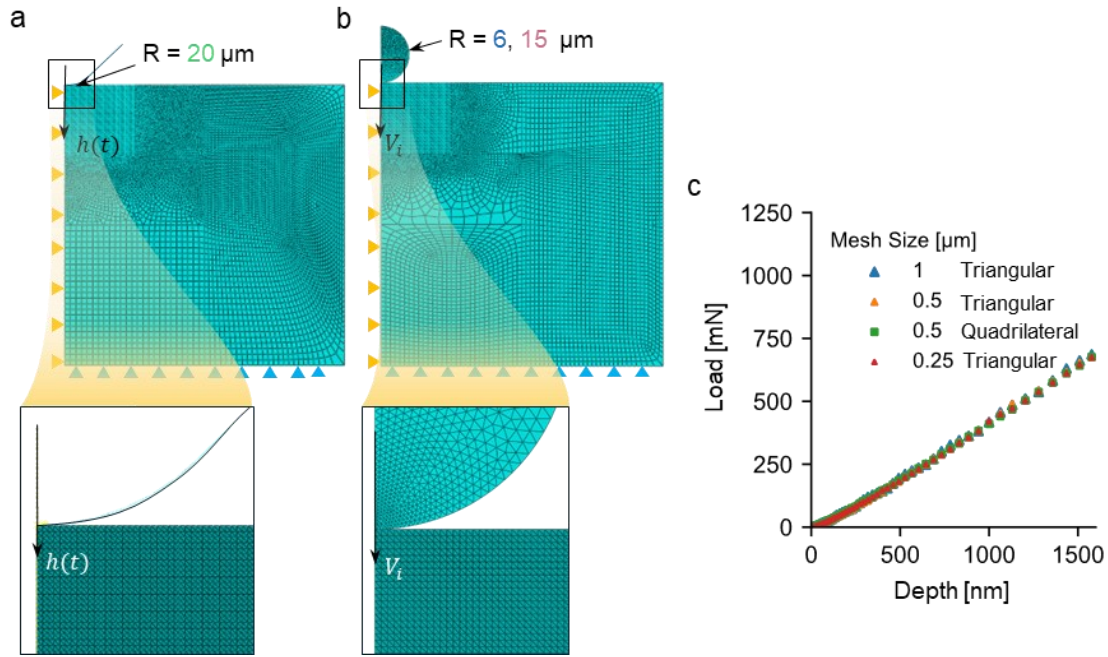

**Fig. S6. Domain discretization.** (a) Mesh geometry used in the finite element simulations of nanoindentation. (b) Mesh geometry for LIPIT impacts with 12 µm- and 30 µm-diameter projectiles. (c) Mesh sensitivity analysis.

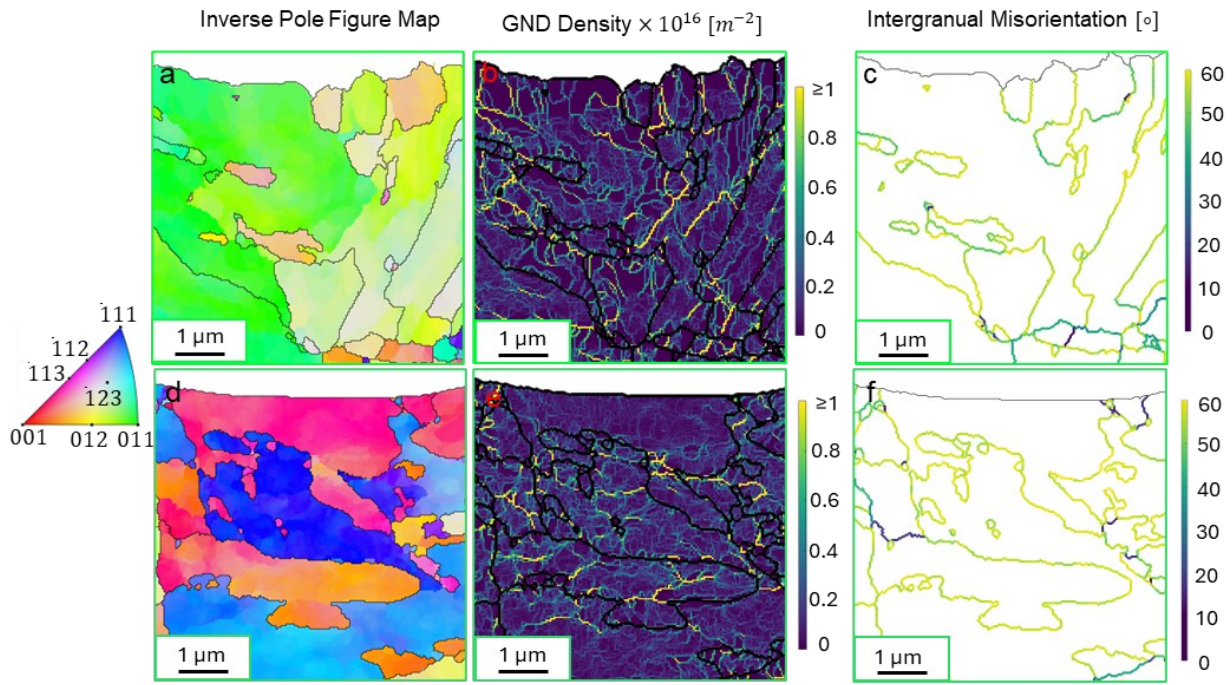

**Fig. S7. Microstructural evolution after nanoindentation**. (a) and (d) EBSD orientation maps taken beneath the impact craters after the NI of the LCS at $0.05\ s^{-1}$ (a) and at $5\ s^{-1}$(d); (b) and (e) the corresponding intragranular geometrically necessary dislocation density and (c) and (f) intergranular misorientation maps.

Table S1. Summary of the re-indentation experiments results

| Test Number | Material | Prior Deformation Strain Rate $\dot{\varepsilon}\ (s^{-1})$ | Young's Modulus E (GPa) | $\sigma_E$ (GPa, propagated) | Hardness H (GPa) | $\sigma_H$ (GPa, propagated) |
|---|---|---|---|---|---|---|
| 1 | LCS | 4.80E+00 | 232.91 | 17.26 | 5.00 | 2.42E-03 |
| 2 | LCS | 4.89E+00 | 273.45 | 23.31 | 5.49 | 2.62E-03 |
| 3 | LCS | 4.90E+00 | 217.83 | 27.73 | 5.44 | 2.40E-03 |
| 4 | LCS | 4.85E-01 | 169.46 | 11.59 | 4.25 | 2.00E-03 |
| 5 | LCS | 4.91E-01 | 206.43 | 17.05 | 4.30 | 2.27E-03 |
| 6 | LCS | 4.81E-01 | 227.38 | 15.07 | 6.17 | 2.75E-03 |
| 7 | LCS | 4.92E-02 | 198.13 | 13.51 | 4.44 | 1.65E-03 |
| 8 | LCS | 4.95E-02 | 248.03 | 26.38 | 4.47 | 1.74E-03 |
| 9 | LCS | 5.06E+01 | 150.20 | 13.03 | 5.08 | 2.31E-03 |
| 10 | LCS | 5.12E+01 | 271.80 | 26.75 | 5.25 | 2.98E-03 |
| 11 | LCS | 5.13E+01 | 141.24 | 12.64 | 4.67 | 1.34E-03 |
| 12 | LCS | 2.72E+02 | 205.23 | 17.76 | 4.85 | 2.01E-03 |
| 13 | LCS | 2.77E+02 | 182.73 | 14.92 | 4.40 | 1.81E-03 |
| 14 | LCS | 8.80E+06 | 203.06 | 12.94 | 5.41 | 3.22E-03 |
| 15 | LCS | 1.02E+07 | 245.41 | 27.76 | 6.04 | 4.50E-03 |
| 16 | LCS | 9.97E+06 | 220.15 | 22.71 | 4.95 | 1.95E-03 |
| 17 | LCS | 7.17E+06 | 241.44 | 16.94 | 7.06 | 6.15E-03 |
| 18 | LCS | 8.90E+06 | 179.85 | 16.60 | 4.85 | 3.08E-03 |
| 19 | LCS | 1.03E+07 | 240.15 | 20.73 | 6.04 | 4.45E-03 |
| 20 | LCS | 3.08E+07 | 223.07 | 26.80 | 5.64 | 1.01E-02 |
| 21 | LCS | 3.84E+07 | 231.73 | 15.27 | 6.80 | 9.83E-03 |
| 22 | LCS | 2.41E+07 | 214.61 | 22.33 | 4.99 | 5.68E-03 |
| 23 | LCS | 3.71E+07 | 183.20 | 14.58 | 6.05 | 3.77E-03 |
| 24 | LCS | 3.75E+07 | 234.87 | 18.41 | 6.41 | 1.85E-03 |
| 25 | LCS | 4.17E+07 | 335.95 | 31.84 | 6.19 | 2.67E-03 |
| 26 | LCS | 3.34E+07 | 215.05 | 15.04 | 5.93 | 3.96E-03 |
| 27 | Pure Iron | 4.15E+07 | 183.89 | 11.93 | 3.45 | 5.06E-03 |
| 28 | Pure Iron | 3.63E+07 | 241.51 | 26.41 | 4.36 | 5.04E-03 |
| 29 | Pure Iron | 3.25E+07 | 192.76 | 17.42 | 3.65 | 3.80E-03 |
| 30 | Pure Iron | 3.13E+07 | 143.64 | 17.57 | 3.38 | 2.44E-03 |
| 31 | Pure Iron | 3.92E+07 | 187.98 | 23.26 | 2.99 | 2.62E-03 |

**References**

[1] F.M. White, Fluid mechanics, 8. ed, McGraw-Hill, New York, NY, 2016.
[2] J. Almedeij, Drag coefficient of flow around a sphere: Matching asymptotically the wide trend, Powder Technology 186 (2008) 218–223. https://doi.org/10.1016/j.powtec.2007.12.006.

[3] Håkon Wiik Ånes, Lars Andreas Hastad Lervik, Ole Natlandsmyr, Tina Bergh, Eric Prestat, Andreas V. Bugten, Erlend Mikkelsen Østvold, Zhou Xu, Carter Francis, Magnus Nord, pyxem/kikuchipy: kikuchipy 0.11.2, (2025). https://doi.org/10.5281/ZENODO.3597646.

[4] M.A. Jackson, E. Pascal, M. De Graef, Dictionary Indexing of Electron Back-Scatter Diffraction Patterns: a Hands-On Tutorial, Integr Mater Manuf Innov 8 (2019) 226–246. https://doi.org/10.1007/s40192-019-00137-4.

[5] M. Hassani, D. Veysset, K.A. Nelson, C.A. Schuh, Material hardness at strain rates beyond $10^6$ s$^{-1}$ via high velocity microparticle impact indentation, Scripta Materialia 177 (2020) 198–202. https://doi.org/10.1016/j.scriptamat.2019.10.032.

[6] W. Pantleon, Resolving the geometrically necessary dislocation content by conventional electron backscattering diffraction, Scripta Materialia 58 (2008) 994–997. https://doi.org/10.1016/j.scriptamat.2008.01.050.

[7] E. Demir, D. Raabe, N. Zaafarani, S. Zaefferer, Investigation of the indentation size effect through the measurement of the geometrically necessary dislocations beneath small indents of different depths using EBSD tomography, Acta Materialia 57 (2009) 559–569. https://doi.org/10.1016/j.actamat.2008.09.039.

[8] P.J. Konijnenberg, S. Zaefferer, D. Raabe, Assessment of geometrically necessary dislocation levels derived by 3D EBSD, Acta Materialia 99 (2015) 402–414. https://doi.org/10.1016/j.actamat.2015.06.051.

[9] A. Vilalta-Clemente, G. Naresh-Kumar, M. Nouf-Allehiani, P. Gamarra, M.A. Di Forte-Poisson, C. Trager-Cowan, A.J. Wilkinson, Cross-correlation based high resolution electron backscatter diffraction and electron channelling contrast imaging for strain mapping and dislocation distributions in InAlN thin films, Acta Materialia 125 (2017) 125–135. https://doi.org/10.1016/j.actamat.2016.11.039.

[10] A.J. Wilkinson, G. Meaden, D.J. Dingley, High resolution mapping of strains and rotations using electron backscatter diffraction, Materials Science and Technology 22 (2006) 1271–1278. https://doi.org/10.1179/174328406X130966.

[11] F. Bachmann, R. Hielscher, H. Schaeben, Texture Analysis with MTEX – Free and Open Source Software Toolbox, SSP 160 (2010) 63–68. https://doi.org/10.4028/www.scientific.net/SSP.160.63.

[12] L. Chamma, J.-M. Pipard, A. Arlazarov, T. Richeton, S. Berbenni, A plasticity-induced internal length mean field model based on statistical analyses of EBSD and nanoindentation data, International Journal of Plasticity 189 (2025) 104327. https://doi.org/10.1016/j.ijplas.2025.104327.

[13] R. Fernández, G. Bokuchava, G. Bruno, I. Serrano-Muñoz, G. González-Doncel, On the dependence of creep-induced dislocation configurations on crystallographic orientation in pure Al and Al–Mg, J Appl Crystallogr 56 (2023) 764–775. https://doi.org/10.1107/S1600576723003771.

[14] J.W. Kysar, Y. Saito, M.S. Oztop, D. Lee, W.T. Huh, Experimental lower bounds on geometrically necessary dislocation density, International Journal of Plasticity 26 (2010) 1097–1123. https://doi.org/10.1016/j.ijplas.2010.03.009.

[15] A.C. Leff, C.R. Weinberger, M.L. Taheri, Estimation of dislocation density from precession electron diffraction data using the Nye tensor, Ultramicroscopy 153 (2015) 9–21. https://doi.org/10.1016/j.ultramic.2015.02.002.